\documentclass[fleqn,usenatbib]{mnras}
\usepackage{newtxtext,newtxmath}
\usepackage[T1]{fontenc}

\DeclareRobustCommand{\VAN}[3]{#2}
\let\VANthebibliography\thebibliography
\def\thebibliography{\DeclareRobustCommand{\VAN}[3]{##3}\VANthebibliography}

\usepackage{amsmath}	
\usepackage{graphicx}	
\usepackage{xspace}	    

\newcommand{\Msun}{\mathrm{M}_\odot}

\newcommand{\HII}{{H\textsc{ii}}\xspace}
\newcommand{\sC}{{$^\mathrm{s}$C}}
\newcommand{\sK}{{$^\mathrm{s}$K}}
\newcommand{\dC}{{$^\mathrm{d}$C}}
\newcommand{\dT}{{$^\mathrm{d}$T}}

\title[Synthetic UV-submm images for ARTEMIS]{High-resolution synthetic UV–submm images for Milky Way-mass simulated galaxies from the ARTEMIS project}

\author[Camps et al.]{
Peter Camps,$^{1}$\thanks{E-mail: peter.camps@ugent.be}
Anand Utsav Kapoor,$^{1}$
Ana Trčka,$^{1}$
Andreea S. Font,$^{2}$
Ian G. McCarthy,$^{2}$
James Trayford,$^{3}$
\newauthor
and Maarten Baes$^{1}$
\\
$^{1}$Sterrenkundig Observatorium, Universiteit Gent, Krijgslaan 281, B-9000 Gent, Belgium
\\
$^{2}$Astrophysics Research Institute, Liverpool John Moores University, 146 Brownlow Hill, Liverpool L53RF, UK
\\
$^{3}$Institute of Cosmology and Gravitation, University of Portsmouth, Dennis Sciama Building, Burnaby Road, Portsmouth PO1 3FX, UK
}

\pubyear{2021}

\begin{document}
\label{firstpage}
\pagerange{\pageref{firstpage}--\pageref{lastpage}}
\maketitle

\begin{abstract}
We present redshift-zero synthetic dust-aware observations for the 45 Milky Way-mass simulated galaxies of the ARTEMIS project, calculated with the SKIRT radiative transfer code. The post-processing procedure includes components for star-forming regions, stellar sources, and diffuse dust. We produce and publicly release realistic high-resolution images for 50 commonly-used broadband filters from ultraviolet to sub-millimetre wavelengths and for 18 different viewing angles. We compare the simulated ARTEMIS galaxies to observed galaxies in the DustPedia database with similar stellar mass and star formation rate, and to synthetic observations of the simulated galaxies of the Auriga project produced in previous work using a similar post-processing technique. In all cases, global galaxy properties are derived using SED fitting. We find that, similar to Auriga, the post-processed ARTEMIS galaxies generally reproduce the observed scaling relations for global fluxes and physical properties, although dust extinction at FUV/UV wavelengths is underestimated and representative dust temperatures are lower than observed. At a resolved scale, we compare multi-wavelength non-parametric morphological properties of selected disc galaxies across the data sets. We find that the ARTEMIS galaxies largely reproduce the observed morphological trends as a function of wavelength, although they appear to be more clumpy and less symmetrical than observed. We note that the ARTEMIS and Auriga galaxies occupy adjacent regions in the specific star formation versus stellar mass plane, so that the synthetic observation data sets supplement each other.
\end{abstract}



\section{Introduction}

The plethora of data produced by current and planned earth-based observatories and space missions enable an exceedingly detailed study of cosmic structure, and in particular of the assembly and evolution of galaxies. One important method that helps us uncover and make sense of the physical mechanisms underlying galaxy formation is to emulate those processes in computer simulations. The most comprehensive simulations evolve dark and baryonic matter in a cosmologically relevant volume from initial conditions at high redshift to the present day \citep[for a review, see][]{Vogelsberger2020a}, employing subgrid recipes for unresolved processes such as star formation, stellar feedback, and chemical evolution. Recent examples include EAGLE \citep{Crain2015,Schaye2015}, MassiveBlack-II \citep{Khandai2015}, Romulus25 \citep{Tremmel2017}, SIMBA \citep{Dave2019}, and Illustris-TNG50 \citep{Pillepich2019,Nelson2019}. These simulations succeed in reproducing many observed global galaxy properties to a fair degree, including for example stellar mass functions, galaxy sizes, mass-metallicity relations, star formation relations, passive fractions, and gas contents (see the references listed above for each simulation project).

The resolution of cosmological simulations is necessarily limited by the available computational resources. Smaller simulation volumes allow a somewhat better resolution but reproduce fewer massive structures and rare objects. Using an alternate approach, cosmological zoom simulations focus on a limited portion of a larger simulation volume, for example the contents of a single dark matter halo, to achieve baryon mass resolutions down to around $10^4~\Msun$. Recent examples include NIHAO \citep{Wang2015}, APOSTLE \citep{Sawala2016}, Latte \citep{Wetzel2016}, Auriga \citep{Grand2017}, RomulusC \citep{Tremmel2019}, and ARTEMIS \citep{Font2020,Font2021}. Each resolution element now has a mass at the upper end of the observed molecular cloud mass range, implying that further resolution improvements will likely need to be accompanied by enhanced subgrid recipes to better capture the physical processes on these smaller scales.

To increase the accuracy of future simulation efforts, and thus improve our understanding of the emulated physics, a detailed comparison of simulation results to observations is required. Comparing simulation results to observations, however, is often tricky. Simulations yield intrinsic galaxy properties such as mass, age, metallicity, or star formation rate (SFR) by aggregating the corresponding properties of the particles or cells used to represent physical constituents. Observations, on the other hand, yield multi-wavelength fluxes and spectra which are integrated along the line of sight and thus provide a two-dimensional projection of the galaxy under study. The observed radiation is often significantly altered by the effects of dust grains in the interstellar medium (ISM) \citep[e.g.,][]{Viaene2016} and depends non-linearly on the complex geometry of the galaxy \citep[e.g.,][]{Saftly2015}. As a result, deriving intrinsic properties from the observed data always involves some form of conversion that relies on approximating assumptions \citep[e.g.,][]{Kennicutt2012,Courteau2014}.

Alternatively, one can bring the simulation output into the observational realm through forward modelling. In addition to assigning appropriate emission spectra, this requires simulating transport of the radiation through the ISM, including the scattering, absorption and re-emission by dust grains. This can be accomplished using radiative transfer (RT) codes such as Grasil3D \citep{DominguezTenreiro2014}, Hyperion/Powderday \citep{Robitaille2011, Narayanan2021}, or SKIRT \citep{Baes2011, Camps2015a, Camps2020}. While this approach obviously also relies on approximations, it offers important benefits. It allows incorporation of a wide range of physics into the model, including for example the detailed distribution of stars and dust in the simulated galaxy. The synthetic observables resulting from the model can be directly compared to observed data and can be processed or visualised using any of the tools commonly used to interpret observations. This includes deriving estimates for the physical properties from the synthetic observations, which can help evaluate the employed recipes by comparison to the known intrinsic properties of the simulated galaxies.

In the past decade, many authors have taken this route to generate synthetic observables for cosmological (zoom) simulations \citep[e.g.,][]{Jonsson2010, Lanz2014, Granato2015, Bignone2016, Camps2016, Trayford2017, SantosSantos2017, Lahen2018, Gjergo2018, Barber2018, Narayanan2018, RodriguezGomez2019, Ma2019, Cochrane2019, Liang2019, Vogelsberger2020b, Parsotan2021, Lovell2021, Granato2021}. Some also prepare data sets for public use. For example, \citet{Camps2018a} publish spatially integrated UV to submm broadband fluxes for nearly half a million of EAGLE galaxies up to redshift 6. More recently, \citet{Trcka2021} offer a similar data set for the Illustris-TNG50 galaxies, and \citet{Kapoor2021} provide high-resolution broadband images for 30 present-day Auriga zoom galaxies.

In this work we consider the recent ARTEMIS project \citep{Font2020,Font2021}, encompassing 45 zoom simulations of Milky Way-mass dark matter mass haloes performed using the EAGLE cosmological simulation code \citep{Crain2015,Schaye2015}. We use SKIRT version 9 \citep{Camps2020} to produce synthetic observations at redshift zero for the main galaxy in each of the haloes, with a stellar mass ranging from $1$ to $9\times10^{10}~\Msun$. We calibrate our RT post-processing scheme by comparing to observed galaxies with similar stellar mass and SFR from the DustPedia data set \citep{Clark2018}, a large sample of nearby galaxies with matched aperture photometry in more than 40 bands from UV to millimetre wavelengths. For each ARTEMIS galaxy, we publish highly resolved images ($50\times50$~pc pixels) observed for 18 sight lines through 50 commonly used broadband filters spanning ultraviolet (UV) to sub-millimetre (submm) wavelengths.

Our setup and procedures are nearly identical to those employed by \citet{Kapoor2021} for producing synthetic observations of the galaxies in the Auriga project \citep{Grand2017}. The latter project comprises 30 zoom simulations of isolated Milky Way-mass dark matter haloes, selected from a dark-matter-only simulation and evolved to redshift zero in a full cosmological context including baryon physics. The main galaxy in each halo has a stellar mass in a range of $3$ to $11\times10^{10}~\Msun$. Although the Auriga and ARTEMIS mass ranges are similar, the Auriga galaxies are more massive on average and nearly all of them are spiral galaxies because of the employed selection criteria. The ARTEMIS galaxies show a much more diverse morphology and occupy a different region in the specific star formation rate (sSFR) versus stellar mass plane.

As a result, the data set for ARTEMIS prepared in this work complements and augments the data set prepared by \citet{Kapoor2021} for Auriga in several ways. The number of simulated galaxies for which high-resolution and multi-wavelength synthetic observables are made available is more than doubled, from 30 to 75. The stellar mass range is extended downwards and more diverse morphology types are included, so that the combined data set allows studying simulated galaxy properties, scaling relations and dust heating on resolved scales for a wide range of Milky Way-mass galaxies. It thus becomes possible to compare spatially resolved properties of the ARTEMIS and the Auriga galaxies with observations and among each other, leading to a better understanding of how well each simulation reproduces reality and why.

In Sect.~\ref{sec:background} we briefly describe the cosmological simulations, the observed data, and the software codes used in this work. In Sect.~\ref{sec:methods} we review our RT post-processing procedure, discuss the calibration of the associated parameters, and list the data products being made available as a result. In Sect.~\ref{sec:analysis} we then offer an initial analysis of these synthetic observables, including global physical properties derived by SED fitting using CIGALE \citep{Boquien2019} and non-parametric morphology properties calculated from the spatially resolved images at several wavelengths using StatMorph \citep{RodriguezGomez2019}. We compare these properties to those of similar observed DustPedia galaxies and simulated Auriga galaxies, and discuss the implications. In Sect.~\ref{sec:conclusions} we summarise and conclude.


\section{Background}
\label{sec:background}

\subsection{ARTEMIS}
\label{sec:artemis}

The ARTEMIS project includes a set of 45 zoomed-in, high-resolution hydrodynamical simulations of galaxies residing in haloes of Milky Way mass, 42 of them presented by \citet{Font2020} and 3 more by \citet{Font2021}. The baryon mass resolution is about $3\times10^4~\Msun$. The simulations are performed with the EAGLE galaxy formation code \citep{Crain2015,Schaye2015} using the same solvers and subgrid physics except for a re-calibrated stellar feedback recipe. The simulation setup is fully described by \citet{Font2020} and references therein; we provide just a brief summary here.

The MUSIC code \citep{Hahn2011} is used to generate initial conditions at redshift 127 for a flat $\Lambda$CDM WMAP9 cosmology \citep{Hinshaw2013} in a periodic box 25 Mpc on a side. This volume is then evolved to redshift zero using collisionless dynamics. From the completed simulation a volume-limited sample is selected of all 63 haloes within a mass range of $8\times10^{11} < M_{200}/\Msun < 2\times10^{12}$, where $M_{200}$ is the mass enclosing a mean density of 200 times the critical density at redshift zero. The selection is based solely on halo mass with no conditions on the merger history or environment.

For 45 of the selected haloes, hydrodynamic zoom simulations are performed using full baryonic physics at high resolution within a region enclosing twice the halo radius, and using dark-matter-only dynamics at lower resolution in the remainder of the volume. The EAGLE code employed to run the zoom simulations is a modified version of the N-body smoothed particle hydrodynamics (SPH) code GADGET-3 \citep{Springel2005}. It provides subgrid models of important processes that cannot be resolved directly in the simulations, including metal-dependent radiative cooling, star formation, stellar evolution and chemodynamics, black hole formation and growth, stellar feedback, and active galactic nucleus (AGN) feedback. Star formation assumes the \citet{Chabrier2003} initial mass function (IMF).

The efficiency of the stellar feedback in the main EAGLE runs presented in \cite{Schaye2015} was fine-tuned to approximately reproduce the local galaxy stellar mass function and the size-stellar mass relation of disc galaxies. However, with increased numerical resolution, the efficiency of stellar feedback needs to be re-adjusted to recover the good match to the calibration observables. Because the resolution of the ARTEMIS simulations is about 7 times better than the finest mass resolution in the main EAGLE runs, stellar feedback efficiency was recalibrated by increasing the value of the density where the efficiency of stellar feedback transitions to its maximal value. In the EAGLE stellar feedback model, the fraction of available stellar energy used for feedback is modelled with a sigmoid function of density (and metallicity). This function asymptotes to fixed values at low and high densities, such that a higher fraction of the available energy is used at high densities in order to offset spurious (numerical) radiative cooling losses. As we increase the resolution of the simulations, the density scale at which numerical losses become important increases, motivating an increase in the transition density scale used for stellar feedback in ARTEMIS. The transition density scale was adjusted by hand so that the simulations reproduce the amplitude of the stellar mass -- halo mass relation at a halo mass scale of about $10^{12}~\Msun$ \citep[see Fig.~2 of][]{Font2020}. In addition, the observed sizes and star formation rates of these systems were also reproduced, without any explicit calibration to match those quantities (see the same figure).

In this work, we consider the central galaxy (i.e. the most massive object) in the redshift-zero snapshot for each of the 45 ARTEMIS haloes, excluding any satellites or other secondary objects. Where applicable, we indicate particular galaxies using the same identifiers as introduced by  \citet[][G1--G42]{Font2020} and \citet[][G43--G45]{Font2021}.

\subsection{Auriga}
\label{sec:auriga}

The Auriga project \citep{Grand2017} includes a set of cosmological magneto-hydrodynamical zoom simulations of the formation of galaxies in isolated Milky Way mass dark haloes. The baryon mass resolution for the 30 simulations at the standard (level 4) resolution considered here is about $5\times10^4~\Msun$. The simulation setup is fully described by \citet{Grand2017} and references therein; we provide just a brief summary here.

The starting point for the zoom simulations is a dark-matter-only counterpart to the 100~Mpc-box Eagle simulation (L100N1504) introduced in \citep{Schaye2015} and adopting $\Lambda$CDM cosmological parameters taken from \citet{Planck2014}. This parent simulation is evolved from redshift 127 to the present day. The linear phases for the parent simulation, and for all of the zoom simulations, are taken from the public Gaussian white noise field realisation, PANPHASIA \citep{Jenkins2013}. 

Host haloes are selected from the parent simulation through a mass cut criterion of $1\times10^{12} < M_{200}/\Msun < 2\times10^{12}$ and the requirement that each candidate halo be relatively isolated at redshift zero. The degree of isolation is estimated, roughly speaking, by the distance to other haloes in the simulation relative to the virial radius of each candidate halo. From the total of 697 haloes in the chosen mass range, 174 are in the most isolated quartile, and 30 of those are randomly selected for re-simulation.

The zoom simulations are performed with the N-body, magneto-hydrodynamics (MHD) moving mesh code AREPO \citep{Springel2010}, equipped with a comprehensive physics model containing subgrid recipes for processes that cannot be resolved. These recipes are similar to those employed in ARTEMIS, also assuming the \citet{Chabrier2003} IMF, but differ in many details.
We summarise the more relevant ones. (1) The cold gas is not modelled in either simulation. To prevent spurious fragmentation, ARTEMIS imposes a temperature floor corresponding to the equation of state $P \propto \rho^{4/3}$. Auriga implements the two phase model introduced by \citet{Hernquist2003}. (2) ARTEMIS uses a metallicity-dependent star formation threshold, while Auriga employs a fixed threshold. (3) ARTEMIS implements stochastic thermal feedback from core-collapse supernovae; the feedback efficiency is mediated using metallicity and density-dependent factors. Auriga implements core-collapse supernovae feedback by launching wind particles that travel away from the originating site with a given velocity. The energy is deposited once certain criteria are met \citep{Marinacci2015}. (4) ARTEMIS provides a single mode of AGN feedback with a fixed efficiency. The energy is injected thermally at the location of the black hole at a rate that is proportional to the gas accretion rate. This is similar to Auriga's quasar mode. Auriga includes separate quasar and radio modes. For the quasar mode, the thermal energy is injected isotropically into neighboring gas cells. For the radio mode, bubbles of gas are gently heated at randomly placed at locations following an inverse square distance profile around the black hole.

In addition, the Auriga simulations include prescriptions for magnetic field evolution.
In the halo investigated by \citet{vandeVoort2021}, the central galaxy is more disc-dominated and the central black hole is more massive when magnetic fields are included. Also, the physical properties of the circumgalactic medium (CGM) change significantly. On the other hand, the global galaxy properties including stellar mass and SFR remain essentially unaffected.

In this work, we indirectly use the Auriga simulation results through the synthetic observations prepared by \citet{Kapoor2021} for the main galaxy in each of the 30 haloes.

\subsection{DustPedia}

The DustPedia project \citep{Davies2017} combines observations from the \emph{Herschel} and \emph{Planck} missions and several other sources to study dust and dust-related processes in local galaxies. One outcome of the project is a public data set providing matched aperture photometry in more than 40 bands from UV to millimeter wavelengths for a sample of 875 nearby galaxies at distances up to $\approx$40~Mpc \citep{Clark2018}.

\citet{Casasola2020} study ISM scaling relations for the galaxies in the DustPedia data set. They report that the selection and uniform treatment of the DustPedia data leads to a complete and homogeneous galaxy sample covering a broad dynamic range of various physical properties, including stellar mass, SFR, and morphological stage. This makes the DustPedia data set ideally suited to put constraints on cosmological simulations predicting ISM properties and scaling relations. For example, \citet{Trcka2020} compare the global properties and scaling relations of galaxies produced by the EAGLE simulations to those observed for the DustPedia galaxies. 

\citet{Kapoor2021} use the DustPedia data set to calibrate their RT post-processing recipe for the simulated Auriga galaxies, and subsequently compare selected scaling relations and morphological properties between simulations and observations. In this work, we follow in their path for the RT post-processing of the ARTEMIS galaxies. 

\subsection{SKIRT}

The SKIRT code\footnote{The open-source SKIRT code is registered at the ASCL with the code entry ascl:1109.003. Documentation and other information can be found at \url{www.skirt.ugent.be}.} \citep{Baes2011, Camps2015a, Camps2020} is a fully three-dimensional Monte Carlo dust RT code equipped with a library of flexible input models \citep{Baes2015}, routines to import the output from various kinds of hydrodynamical simulations \citep{Camps2015a}, a module handling stochastic heating and emission of dust grains \citep{Camps2015b}, and a hybrid parallelization strategy \citep{Verstocken2017, Camps2020}. A range of advanced spatial grids for discretizing the medium is implemented in SKIRT, including methods to efficiently traverse photons through these grids \citep{Camps2013, Saftly2013, Saftly2014}.

SKIRT has been extensively used to generate synthetic UV to submm broadband images, spectral energy distributions and polarisation maps for idealised galaxies \citep[e.g.,][]{Baes2003, Gadotti2010, DeGeyter2014, Lee2016, Peest2017}, for high-resolution 3D galaxy models \citep[e.g.,][]{DeLooze2014, Verstocken2020, Nersesian2020a, Nersesian2020b, Viaene2020}, and for galaxies extracted from cosmological simulations \citep[e.g.,][]{Saftly2015, Camps2016, Camps2018a, Trayford2017, Liang2018, Behrens2018, Lahen2018, Barber2018, RodriguezGomez2019, Ma2019, Vogelsberger2020b, Parsotan2021, Granato2021, Kapoor2021}.

In this work we use SKIRT version 9\footnote{Specifically, git commit \texttt{c70b6ef06ca5} in the master branch of the SKIRT code hosted at \url{www.github.com/SKIRT/SKIRT9}} to produce synthetic observations for the ARTEMIS galaxies, after extracting the relevant information from the simulation snapshots through straightforward Python procedures. The full procedure and configuration details are discussed in Sect.~\ref{sec:methods}.

\subsection{CIGALE}
\label{sec:cigale}

The CIGALE SED fitting code \citep{Noll2009, Boquien2019} incorporates stellar, nebular, dust emission and dust attenuation. It contains an implementation of a delayed and truncated star-formation history (SFH) \citep{Ciesla2016}, \citet{Bruzual2003} simple stellar population (SSP) libraries, the modified \citet{Calzetti2000} attenuation law, and several dust models.

In this work we use CIGALE version 0.12.1 to estimate global physical properties such as stellar mass, dust mass and SFR from the available broadband fluxes for the various data sets under study, i.e. the ARTEMIS, Auriga and DustPedia galaxies. This allows us to compare the properties of simulated and observed galaxies on equal footing. We therefore use the same parameter settings in all cases.

Specifically, following \citet{Kapoor2021}, we employ the settings used by \citet{Bianchi2018}, \citet{Nersesian2019} and \citet{Trcka2020}, including the THEMIS dust model \citep{Jones2017} which we also use in our RT post-processing procedure (see Sect.~\ref{sec:methods}), except that we specify the \citet{Chabrier2003} IMF for SSPs, consistent with the IMF used in the ARTEMIS and Auriga simulations. For our analysis, we always use the properties corresponding to the most probable `Bayes' model determined by CIGALE.

\subsection{StatMorph}
\label{sec:statmorph}

StatMorph \citep{RodriguezGomez2019} is a Python package for calculating many commonly used non-parametric morphological statistics of galaxy images, including the Gini-M20 \citep{Lotz2004} and concentration-asymmetry-smoothness \citep[CAS,][]{Conselice2003} statistics, and for fitting 2D Sérsic profiles. The code can handle images with a single source each, which is the mode used in this work, as well as large mosaic images with hundreds or thousands of sources. 

In this work we use the exact same procedure as \citet{Kapoor2021} to obtain multi-wavelength sets of the elliptical half-light radii and of the CAS indices for a subset of ARTEMIS disc galaxies. This allows us to compare these morphological properties with those already calculated by \citet{Kapoor2021} for similarly selected Auriga galaxies and by \citet{Baes2020} for a set of well-resolved DustPedia spiral galaxies.

The four statistics studied in this work are described in detail by \citet{RodriguezGomez2019} and references therein. We limit the discussion here to a very brief summary.

\begin{itemize}
  \item \textit{Half light radius} ($R_\mathrm{half}/R_{80}^\mathrm{opt}$): The half-light radius $R_\mathrm{half}$ is calculated as the elliptical radius of the isophote that contains half of the light in the galaxy image. We normalise it with $R_{80}^\mathrm{opt}$, the radius of a circular aperture containing 80\% of the galaxy’s light in the Sloan Digital Sky Survey (SDSS) $g$ band image.
  \item \textit{Concentration} (C): The concentration index is defined as $5 \times \log_{10}(R_{80}/R_{20})$, where $R_{20}$ and $R_{80}$ are the radii of circular apertures containing $20\%$ and $80\%$ of the galaxy's light, respectively. The index is a measure of how concentrated the central region or bulge is with respect to the total flux of the galaxy.
  \item \textit{Asymmetry} (A): The asymmetry index is obtained by subtracting the galaxy image rotated by $180^\circ$ from the original image. Asymmetry indicates merger events and interactions, or, in regular star-forming galaxies, structures such as spiral arms. 
  \item \textit{Smoothness} (S): The smoothness index is computed by subtracting a lower resolution version of the galaxy image from the original galaxy image. It measures the presence of high spatial frequency features; the index value increases with clumpiness.
\end{itemize}


\section{Methods}
\label{sec:methods}

Our procedure for preparing synthetic observables of the simulated ARTEMIS galaxies closely follows the procedure employed by \citet{Kapoor2021} for the simulated Auriga galaxies, which is in turn based on the procedure employed by \citet{Camps2016} and \cite{Trayford2017} for the simulated EAGLE galaxies. This similarity in approach makes the results comparable and allows the combined Auriga and ARTEMIS results to be considered as a single, consistent data set.

\subsection{Extraction and choice of aperture}
\label{sec:aperture}

For each of the 45 ARTEMIS redshift-zero simulation snapshots, we locate the dominant halo, i.e. the halo with the largest stellar mass in its dominant sub-halo, and extract all star and gas particles from that dominant sub-halo. This represents the galaxy of interest in the zoom-in simulation. We do \emph{not} include the other sub-halo's of the dominant halo, representing secondary structures bound to the main galaxy, because these would only interfere with the analysis of the main galaxy. We then transform the particle coordinates so that the origin is in the stellar centre of mass and the $z$-axis coincides with the stellar rotation axis. As a sanity check, we verify that the intrinsic\footnote{We use the adjective \emph{intrinsic} to indicate a quantity obtained by simply aggregating particle properties.} stellar masses within a spherical aperture of 30~kpc match the stellar masses listed in Table 1 of \citet{Font2020}.

We subsequently preserve only those particles with a position inside a spherical aperture with the largest of the following radii: the 30~kpc radius commonly used for EAGLE \citep{Schaye2015}, the radius at which the face-on stellar surface density within $\pm 10$~kpc of the mid-plane in the vertical direction falls to $2\times 10^5 \Msun \,\mathrm{kpc}^{-2}$ \citep[following][]{Kapoor2021}, and 5 times the half-stellar-mass radius or $5\mathrm{R}_{\mathrm{M}50}$ (for optimal comparison with DustPedia observables; see below). We call the largest of these radii the \emph{extraction} aperture. The spatial domain of the SKIRT simulation and the field of view of the generated images are adjusted for each galaxy to enclose its extraction aperture.

This approach allows calculating spatially integrated fluxes from the generated images for any of the apertures listed above, or in fact for any aperture up to the extraction aperture. However, the surface brightness for each pixel, and thus the spatially integrated fluxes, will always reflect the line-of-sight radiation across the complete extraction volume. In other words, the smaller apertures are circular (or cylindrical) rather than spherical.

Tests reported by \citet{Trcka2021} indicate that the $5\mathrm{R}_{\mathrm{M}50}$ radius offers the best match to the apertures in the DustPedia galaxy sample. Therefore, all spatially integrated quantities shown and discussed in this work are calculated for that aperture. This includes the luminosities used for recipe calibration, although the choice of aperture does not seem to have a significant effect on the comparisons. Nevertheless, producing and publishing images with the full extraction aperture (see Sect.~\ref{sec:datadescription}) enables other studies, such as the morphology calculations in Sect.~\ref{sec:morphology}, to compare with data sets that use a different aperture definition.


\begin{table}
 \centering
 \caption{The optimal global dust-to-metal ratio $f_\mathrm{dust}$ found by \citet{Camps2016} for the EAGLE simulations, by \citet{Kapoor2021} for the Auriga simulations, and in Sect.~\ref{sec:calibration} of this work for the ARTEMIS simulations, for the two diffuse dust allocation recipes described in Sect.~\ref{sec:dustrecipes}.}
 \begin{tabular}{c c c c}
  \hline
  Allocation & $f_\mathrm{dust}$ & $f_\mathrm{dust}$ & $f_\mathrm{dust}$ \\
  recipe & EAGLE & Auriga & ARTEMIS \\
  \hline
  \dC16 & 0.300 & 0.225 & 0.300 \\
  \dT12 &   --  & 0.140 & 0.275 \\
  \hline
 \end{tabular}
 \label{tab:fdust}
\end{table}


\begin{figure}
  \centering
  \includegraphics[width=0.9\columnwidth]{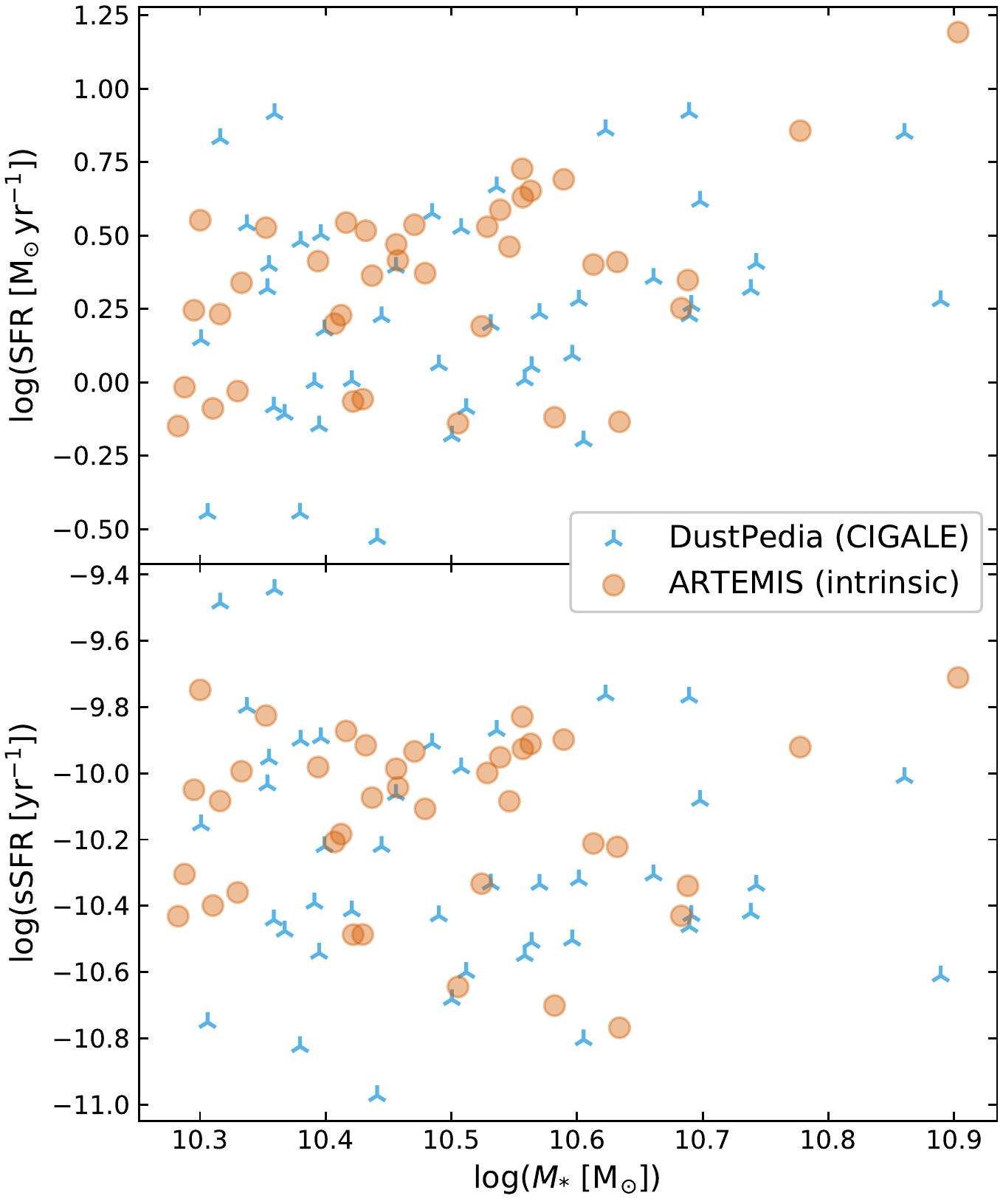}
  \caption{SFR (top) and sSFR (bottom) versus stellar mass for the observed DustPedia galaxies (physical properties obtained through SED fitting) and the simulated ARTEMIS galaxies (intrinsic properties from snapshot particles) used for calibrating our RT post-processing recipe. The selection criteria for the calibration sample are discussed in Sect.~\ref{sec:sampleselection}.}
  \label{fig:CalibrationSample}
\end{figure}


\subsection{Post-processing recipes}
\label{sec:recipes}

During the calibration phase, we explore and fine-tune several variations of our RT post-processing recipe, before finally settling on a single fiducial recipe. We describe the recipe and its variations in this section and report on the calibration results in Sect.~\ref{sec:calibration}.

\subsubsection{Common procedures}
\label{sec:commonrecipes}

In all cases, following \citet{Camps2016,Trayford2017,Trcka2020,Kapoor2021}, regular stellar particles are assigned an SED from the \citet{Bruzual2003} template library based on metallicity and age. Also, star-forming region (SF region) particles (as defined in Sect.~\ref{sec:sfrrecipes} below) are assigned an SED from the MAPPINGS III \citep{Groves2008} template library, which models the dust enveloping the core \HII region in addition to the emission from the young stellar objects. Next to the metallicity, this library requires parameters (ambient pressure, compactness and dust covering fraction) that cannot be directly obtained from the snapshot particle properties and thus require an appropriate heuristic as described in Sect.~\ref{sec:sfrrecipes}.

Following recent work, including e.g. \citet{Nersesian2019, Nersesian2020a, Nersesian2020b, Verstocken2020} for constructing 3D models of nearby face-on galaxies and \citet{Kapoor2021} for post-processing the Auriga galaxies, the diffuse dust in our RT simulations is in all cases represented by the THEMIS dust model \citep{Jones2017} as opposed to the Zubko dust model \citep{Zubko2004} used in earlier work \citep[e.g.,][]{Camps2016,Trayford2017}. Where \citet{Zubko2004} explicitly model polycyclic aromatic hydrocarbon (PAH) molecules next to non-composite graphite and silicate grains, the more recent THEMIS model is based on a mixture of amorphous hydrocarbons and amorphous silicates. Our tests indicate that, compared to the Zubko model, for an otherwise fixed recipe, the THEMIS model reduces the discrepancies between simulation and observation in some wavelength regimes but introduces extra tension in other regimes. We explore these differences in more detail in Appendix~\ref{sec:dustmodels}.

An important part of our post-processing procedure is the transformation from the sets of stellar and gas particles extracted from an ARTEMIS snapshot to three distinct sets of particles presented to the SKIRT code: SF regions, regular stellar particles, and particles representing dust. We implement two distinct recipes for handling SF regions, dubbed \sC16 and \sK21 (see Sect.~\ref{sec:sfrrecipes}), and two distinct recipes for allocating the diffuse dust density distribution, dubbed \dC16 and \dT12 (see Sect~\ref{sec:dustrecipes}). We combine these into three different complete recipes for further calibration: \sC16/\dC16, \sC16/\dT12, and \sK21/\dT12 (see Sect.~\ref{sec:combinedrecipes}).

\subsubsection{Recipes for SF regions (\sC16, \sK21)}
\label{sec:sfrrecipes}

The \sC16 recipe follows the procedure prescribed by \citet{Camps2016, Trayford2017} for the EAGLE simulations; see Fig.~2 of \citet{Camps2016}. Young star particles (up to 100~Myr) and gas particles with a nonzero SFR are placed in a temporary bin of candidate SF region particles. These particles are randomly resampled to sub-particles with smaller masses following a power-law molecular cloud mass function with $M\in[700,10^6]\,\Msun$. The sub-particles also receive a random formation time. Sub-particles that have not yet formed join the gas particle bin (which will later be used to allocate dust); infant sub-particles (up to 10~Myr) go into the SF region particle bin, and the remaining sub-particles join the regular stellar particle bin. The SF region sub-particles are also randomly relocated within a small region, increasing the realism of the images.

As the final step, the \sC16 recipe determines the values of the extra parameters needed for the MAPPINGS III templates assigned to the SF region sub-particles. The ambient pressure and compactness are estimated from snapshot particle properties; the dust covering fraction in the photodissociative region (PDR) is set to the fixed value of $f_\mathrm{PDR}=0.1$.

The \sK21 recipe is the same as the one prescribed by \citet{Kapoor2021} for the Auriga simulations. These authors note that the resolution of modern zoom-in simulations brings the mass of the baryon particles well within the range of the molecular cloud mass function so that it is no longer necessary to resample them to smaller masses, even for SF regions. The \sK21 recipe thus simply moves all infant stellar particles to the SF region bin, without involving the star-forming gas particles. This also avoids the need for `converting' between gas and star particles.

The \sK21 recipe determines the MAPPINGS III template parameter values through an alternate approach. The compactness parameter, which essentially controls the temperature of the dust in the SF region, is randomly assigned from a Gaussian distribution motivated by observations \citep{Utomo2019} and previous studies \citep{Kannan2020}. The ambient pressure, which has only a limited effect on the continuum spectrum, is then derived from the compactness and snapshot particle properties. The dust covering fraction is calculated from the age of the SF region (i.e. the infant stellar particle) assuming a fixed molecular cloud dissipation time of $\tau_\mathrm{clear}=1$~Myr.

\subsubsection{Recipes for allocating dust (\dC16, \dT12)}
\label{sec:dustrecipes}

Both recipes for allocating diffuse dust first determine the subset of gas particles deemed to carry dust, and then assign dust to these particles using a fixed dust-to-metal ratio, i.e. $M_\mathrm{dust} = f_\mathrm{dust} Z M_\mathrm{gas}$. The gas mass $M_\mathrm{gas}$ and the metallicity $Z$ are taken from snapshot particle properties, and the dust fraction $f_\mathrm{dust}$ is set to a fixed, global value. The two recipes differ in the heuristic for selecting `dusty' gas particles and the value of $f_\mathrm{dust}$. The latter should be re-calibrated for each recipe through comparison with observations.

The \dC16 recipe again follows \citet{Camps2016, Trayford2017} and is used by \citet{Kapoor2021} under the name \texttt{recSF8000}. This recipe allocates diffuse dust to all gas particles with a nonzero SFR or a gas temperature under 8000~K (or both).

The \dT12 recipe follows \citet{Torrey2015} and is used by \citet{Kapoor2021} under the name \texttt{recT12}. This recipe allocates dust to gas particles that are considered to be rotationally supported according to their position in temperature-density phase space; see Fig.~2 and Eq.~(3) of \citet{Kapoor2021}. Compared to \dC16, \dT12 assigns dust to a larger subset of gas particles, extending the dust density distribution to larger radii and causing it to be less compact and somewhat less clumpy. To compensate for the larger amount of `dusty' gas, the optimal dust-to-metal ratio will be lower than for the \dC16 recipe.

Table~\ref{tab:fdust} lists the optimal values for $f_\mathrm{dust}$ obtained by \citet{Camps2016} for the EAGLE simulations and by \citet{Kapoor2021} for the Auriga simulations. The last column lists the values obtained in this work for the ARTEMIS simulations as described in Sect.~\ref{sec:calibration}. We will also discuss the similarities and differences in these results at the end of that section.

\subsubsection{Combined recipes}
\label{sec:combinedrecipes}

The procedures for handling SF regions and for handling dust are independent of each other and thus the recipes discussed in Sects.~\ref{sec:sfrrecipes} and \ref{sec:dustrecipes} can be combined at will. We will explore the following combinations:
\begin{itemize}
    \item \sC16/\dC16: resampled SF regions and basic dust allocation.
    \item \sC16/\dT12: resampled SF regions and extended dust allocation.
    \item \sK21/\dT12: plain SF regions and extended dust allocation.
\end{itemize}
We do not explore the fourth possible combination, \sK21/\dC16, because \citet{Kapoor2021} found this combination to be slightly less optimal than \sK21/\dT12 when comparing morphology parameters to observations.


\begin{figure*}
  \centering
  \includegraphics[width=\textwidth]{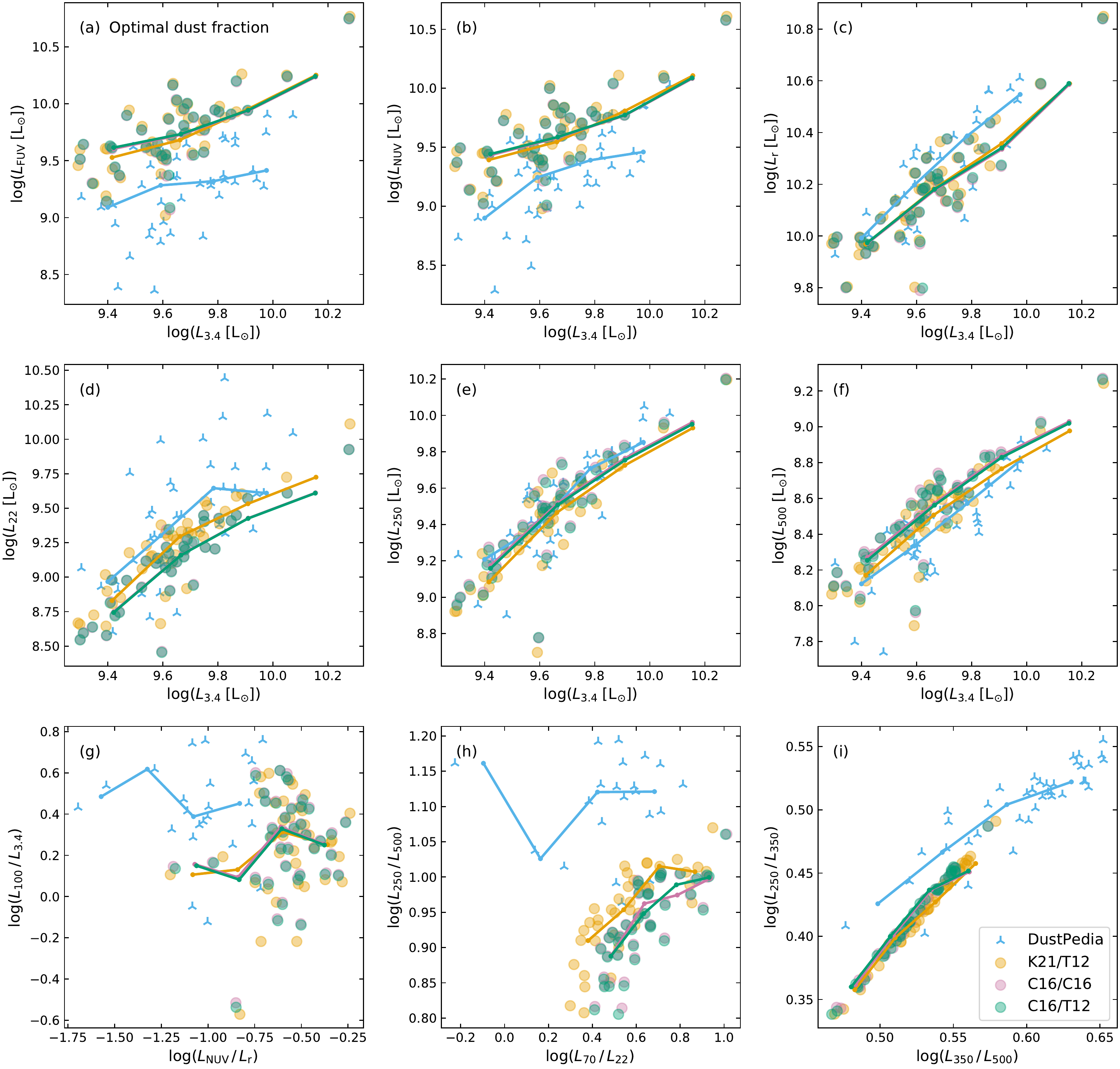}
  \caption{Scaling relations for the synthetic ARTEMIS luminosities calculated for a random viewing angle using our three recipes (Sect.~\ref{sec:combinedrecipes}) \sK21/\dT12 (orange), \sC16/\dC16 (purple), and \sC16/\dT12 (green), each with their optimal dust fraction $f_\mathrm{dust}$ (see Table~\ref{tab:fdust}), versus those for the observed DustPedia luminosities (blue). The solid lines connect the median $y$-axis values in a limited number of $x$-axis bins. The data sets are limited as defined in Sect.~\ref{sec:sampleselection} and, for DustPedia, to the galaxies for which the broadband fluxes under consideration in a given panel have been observed.}
  \label{fig:scalingrelations}
\end{figure*}

\begin{figure*}
  \centering
  \includegraphics[width=0.99\textwidth]{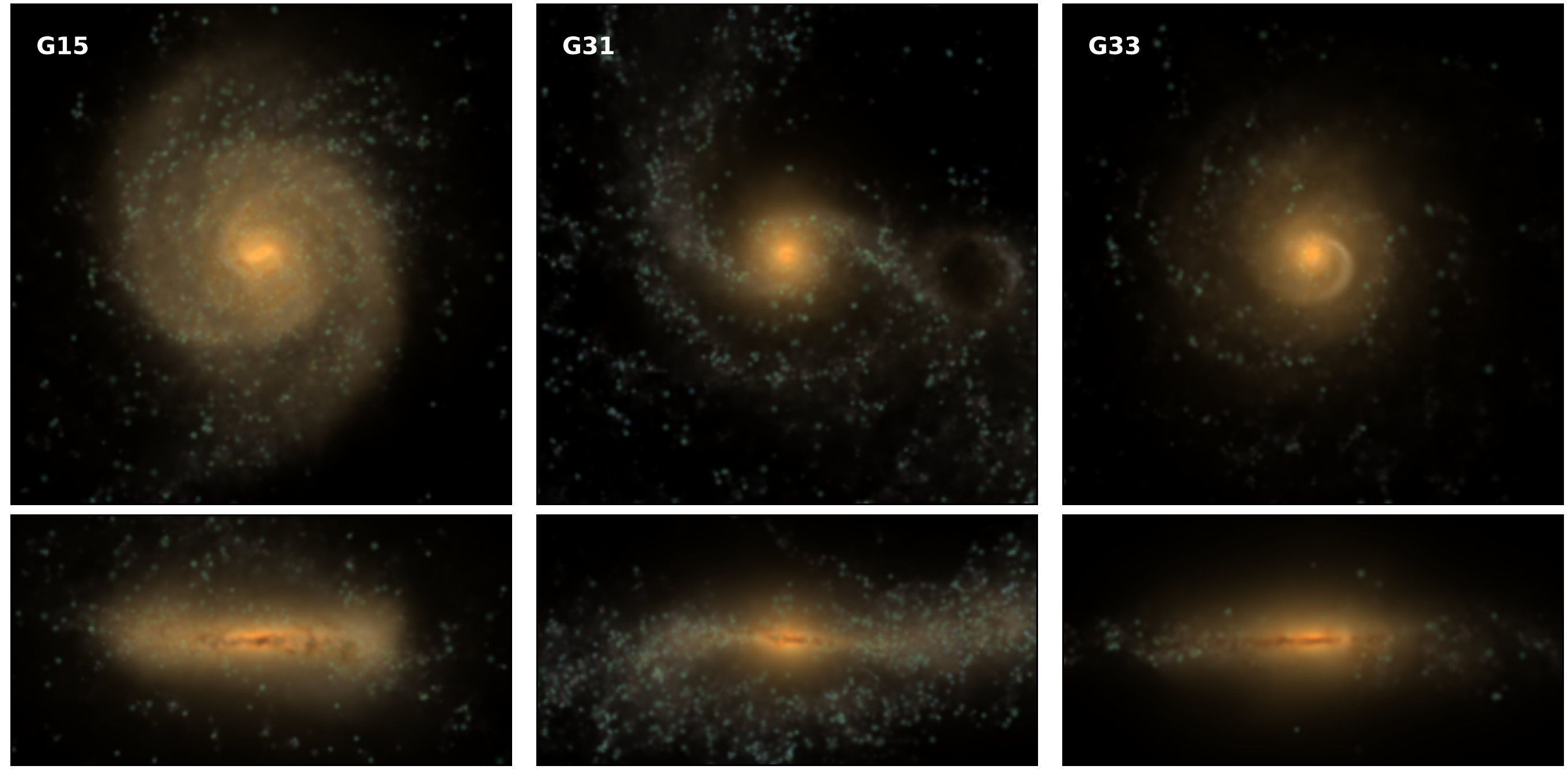}
  \caption{Face-on (top row) and edge-on (bottom row) colour-composite views of selected ARTEMIS galaxies using the synthetic SDSS $i$, $r$, and $g$ observations produced in this work for the red, green, and blue channels, respectively. The field-of view is 60~kpc.}
  \label{fig:RGBImages}
\end{figure*}


\subsection{Calibration}
\label{sec:calibration}

\subsubsection{Sample selection}
\label{sec:sampleselection}

When comparing simulation results with observed data, it is important to approximately match the overall properties of the set of galaxies on both sides. Following \citet{Kapoor2021}, we use stellar mass and SFR criteria for this purpose. For the DustPedia galaxies, we use the physical properties obtained from the observed fluxes through CIGALE SED fitting as described in Sect.~\ref{sec:cigale}. For the ARTEMIS galaxies, we use the intrinsic quantities obtained by accumulating the stellar mass and SFR of the stellar and gas particles, respectively. This is inconsistent in the sense that we are mixing quantities inferred from observed fluxes with intrinsic quantities. Using fluxes for the ARTEMIS galaxies would be circular, however, as we do not yet have a calibrated recipe to produce such fluxes. Moreover, we are using these quantities just to construct approximately matched samples, not to perform the calibration.

It makes no sense to calibrate recipes for handling SF regions and the effects of dust using passive galaxies. We thus eliminate galaxies with $\mathrm{sSFR} < 10^{-11}~\mathrm{yr}^{-1}$ from both the ARTEMIS and DustPedia data sets. Note that we do produce data products for the omitted ARTEMIS galaxies; we just exclude them from the calibration subset. We further limit the DustPedia data set to galaxies with observed stellar mass and SFR in the range of the corresponding intrinsic properties for the ARTEMIS galaxies: $10.30 < \log_{10}(M_*/\Msun) < 10.92$ and $-1.35 < \log_{10}(\mathrm{SFR}/\Msun \, \mathrm{yr}^{-1}) < 1.25$.

Fig.~\ref{fig:CalibrationSample} shows the SFR and sSFR versus stellar mass for the remaining 38 ARTEMIS and 42 DustPedia galaxies. The two populations seem to be similarly distributed and sample the selected parameter space fairly well, demonstrating that the samples are sufficiently comparable for calibrating our post-processing recipe. We note again that Fig.~\ref{fig:CalibrationSample} mixes physical and intrinsic properties. In Sect.~\ref{sec:globalprops} we will investigate how the physical properties derived for the ARTEMIS galaxies through SED fitting relate to their corresponding intrinsic properties.

\subsubsection{Synthetic observations}
\label{sec:calibrationconvergence}

Calibrating our RT post-processing recipe requires performing a significant number of SKIRT simulations for the ARTEMIS galaxies. Therefore, we limit the number of viewing angles and broadbands in this phase. We include an edge-on, a face-on and a \emph{random} view, where the latter corresponds to a sight line looking down from the $z$-axis in the original cosmological coordinate frame (i.e., before the galaxy was rotated). For each of these sight lines, we have SKIRT produce flux densities convolved with the response curve for each of 20 common broadbands ranging from FUV to submm wavelengths and limited to the $5\mathrm{R}_{\mathrm{M}50}$ aperture. We then determine the corresponding absolute luminosities $L = \nu L_\nu = \lambda L_\lambda$ taking into account the configured model-instrument distance.

We subsequently verify numerical convergence of these values with regards to discretization choices made for the simulation, including the resolution of the spatial and spectral grids and the number of photon packets launched, and taking into account the randomness inherent to the Monte Carlo radiative transfer procedure. Our tests confirm that variations in the calculated luminosities caused by numerical issues are always below 8 per cent (0.033 dex) and substantially smaller in most cases.

\subsubsection{Scaling relations}
\label{sec:scalingrelations}

We employ a select set of luminosity scaling relations for comparing the synthetic ARTEMIS results to the observed DustPedia data, as shown in Fig.~\ref{fig:scalingrelations}. The topmost six panels of this figure relate the luminosity for various bands to the 3.4~$\mu$m band luminosity, which can be seen as a proxy for stellar mass. Panels (a), (b), and (c) show the effects of dust attenuation at FUV, NUV, and optical wavelengths, while panels (d), (e), and (f) trace dust emission at infrared and submm wavelengths. The NUV (b) and the 22~$\mu$m (d) luminosities can also be interpreted as a proxy for SFR.

The three panels in the bottom row of Fig.~\ref{fig:scalingrelations} show colour-colour relations in various wavelength regimes. Panel (g) shows a proxy for specific dust mass versus a proxy for specific dust attenuation. Panels (h) and (i) show infrared and submm colour-colour relations indicative of representative dust temperature. In panel (h), a larger contribution of warm dust leads to a lower 70~$\mu$m luminosity relative to 22~$\mu$m and to a lower 500~$\mu$m luminosity relative to 250~$\mu$m. Consequently, data points towards the upper left indicate warmer representative dust temperatures. In panel (i), a similar reasoning leads to the conclusion that data points towards the upper right indicate warmer representative dust temperatures.

In Appendix~\ref{sec:dustfractions} we explore the variations in the ARTEMIS scaling relations for different sight lines and we determine the optimal value of the dust-to-metal fraction $f_\mathrm{dust}$ for each of the recipes \sK21/\dT12, \sC16/\dC16, and \sC16/\dT12 defined in Sect.~\ref{sec:combinedrecipes}. These values are listed in Table~\ref{tab:fdust}. In Fig.~\ref{fig:scalingrelations} and in the discussion here we focus on the ARTEMIS luminosities calculated for a random viewing angle using our three recipes with optimal $f_\mathrm{dust}$.

There is reasonable agreement between the synthetic results and the observed data, with some significant exceptions. In the shorter wavelength regime, the FUV (a) and UV (b) luminosities are overestimated by $\approx 0.5$~dex, while the optical luminosities are underestimated by $\approx 0.15$~dex. These opposing differences cause a correspondingly substantial deviation in the $L_\mathrm{NUV}/L_r$ colour (g). This attenuation discrepancy is in line with the findings of previous work using a similar post-processing recipe \citep[e.g.,][]{Baes2019,Trcka2020,Kapoor2021}. It cannot be resolved by a straightforward scaling of the stellar emission or of the dust mass. It appears that our procedure insufficiently captures the subgrid extinction processes in the compact and clumpy SF regions. In the infrared wavelength regime, the 22~$\mu$m luminosity (d) is underestimated by $\approx 0.25$~dex, depending on the recipe. This is possibly related to the same limitations in our handling of SF regions.

The 100~$\mu$m luminosity (g), more or less at the top of the dust continuum emission peak, is also underestimated by $\approx 0.25$~dex. Luminosities on the long side of the continuum peak (e, f), which may be considered basic proxies for dust mass, seem to be predicted fairly accurately, however with opposing discrepancies along the downward slope. We note $\approx 0.05$~dex underestimation for 250~$\mu$m and up to $\approx 0.15$~dex overestimation for 500~$\mu$m, depending on the recipe. This indicates that the emission peak is shifted to longer wavelengths, corresponding to a larger body of colder dust. This effect is also apparent in the submm-submm colour-colour relation (i) and in the submm-FIR colour-colour relation (h). In both panels, the ARTEMIS data points are significantly shifted toward colder representative dust temperatures compared to the DustPedia observations. Again, as has been noted by \citet{Camps2016} and \citet{Kapoor2021}, our procedure seems to insufficiently capture the subgrid dust heating processes within and in the immediate neighbourhood of the SF regions.

\begin{figure*}
  \centering
  \includegraphics[width=0.8\textwidth]{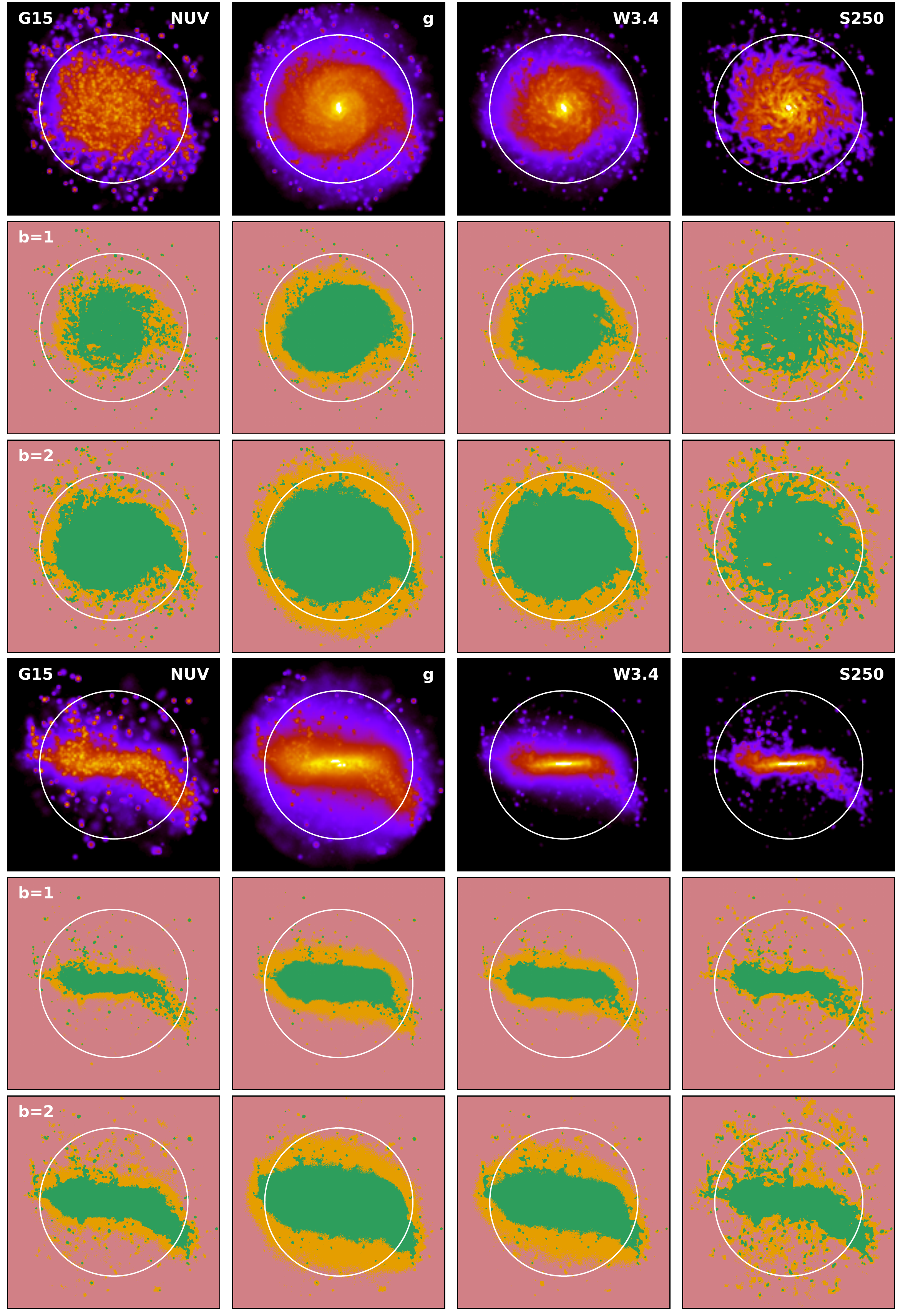}
  \caption{Face-on (top half) and edge-on (bottom half) views of the ARTEMIS galaxy G15 in four bands; from left to right GALEX NUV, SDSS $g$, WISE 3.4~$\mu$m, and \emph{Herschel} SPIRE 250~$\mu$m. G15 has an intrinsic stellar mass of $3.57\times 10^{10}~\Msun$ inside the 30~kpc radius indicated by the white circle. The $5\mathrm{R}_{\mathrm{M}50}$ aperture radius is 36~kpc and the extraction aperture is 43~kpc corresponding to the 86~kpc field of view of the images. The top row in each half shows the synthetic observations produced by SKIRT using a logarithmic colour scale; the transition between red and blue marks 1/100 of the maximum surface brightness. The other two rows in each half show the corresponding convergence statistics indicating reliable (green), questionable (orange) and unreliable (red) image areas assuming 50~pc pixels ($b=1$) and binned 100~pc pixels ($b=2$), as discussed in Sect.~\ref{sec:dataconvergence}.}
  \label{fig:ImageFrames}
\end{figure*}


The differences in the scaling relations for our three recipes are generally small, and most prominent for the longer wavelengths. The \sK21/\dT12 recipe heats the dust somewhat more efficiently than the other recipes (e,f,h,i) and also performs better in the 22~$\mu$m band (d). Both changes can be attributed to the improved handling of SF regions in the \sK21 scheme. Using the metrics discussed in Appendix~\ref{sec:dustfractions} to evaluate the three recipes, the \sK21/\dT12 recipe also robustly emerges as the best recipe. We therefore use this recipe for calculating the final data products of this work.

\subsection{Synthetic data products}
\label{sec:dataproducts}

\subsubsection{Description}
\label{sec:datadescription}

We use the \sK21/\dT12 recipe (see Sect.~\ref{sec:combinedrecipes}) with optimal dust fraction $f_\mathrm{dust}=0.275$ (see Appendix~\ref{sec:dustfractions}) to produce broadband images with a spatial resolution of $50\times50$~pc per pixel for the 45 ARTEMIS galaxies, for 50 bands ranging from UV to submm wavelengths, and for 18 sight lines with varying inclination and azimuth. The lists of broadbands and sight lines match those of the synthetic Auriga observables produced by \citet{Kapoor2021}.

Specifically, we include the 50 broadbands listed in Table~4 of \citet{Camps2018a}, including transmission curves for the most commonly used instruments and observatories across the UV-submm wavelength range. Following the specifications in Sect.~4 of \citet{Kapoor2021}, we use 11 inclinations uniformly sampled in $\cos i$, with $i$ the angle between the angular momentum vector of the galaxy and the line of sight, leading to a finer grid close to edge-on positions. For the three inclinations closest to edge-on, we place observers at three different azimuths. For the remaining eight inclinations, we use just a single azimuthal position. We also consider an additional `random' viewpoint corresponding to a sight line looking down from the $z$-axis in the original cosmological coordinate frame (i.e., before the galaxy was rotated). 

All observers are placed at a distance of 20~Mpc from the simulated galaxy. To determine the field of view (in galaxy size units as opposed to angular units) of the images for a given galaxy, we use the extraction aperture defined in Sect.~\ref{sec:aperture}, which encloses all stellar and gas particles extracted from the corresponding ARTEMIS simulation snapshot. More precisely, the field of view in each image direction is given by twice the extraction aperture radius, rounded up to $64\times50~\mathrm{pc}=3.2~\mathrm{kpc}$. This choice ensures that both the $5\mathrm{R}_{\mathrm{M}50}$ aperture used to calibrate our results against DustPedia observations in Sect.~\ref{sec:calibration} and the surface density-based aperture used by \citet{Kapoor2021} are covered by each image. Furthermore, the rounding  ensures that the number of pixels in each direction is always a multiple of 64, which may facilitate binning of the images in later processing steps. 

These image data are available for public download at \url{https://www.astro.ljmu.ac.uk/Artemis}. As an illustration, Fig.~\ref{fig:RGBImages} offers face-on and edge-on optical views for selected ARTEMIS galaxies, composited from the public data set.

\subsubsection{Convergence}
\label{sec:dataconvergence}

To avoid artefacts caused by the simulation dust grid and limit the noise level in the image pixels, we substantially increase the grid resolution and the number of photon packets compared to the calibration simulation parameters discussed in Sect.~\ref{sec:calibrationconvergence}. Depending on the dust distribution in the galaxy, the number of spatial grid cells ranges from 2 to 13 million, with the cells in the densest diffuse dust regions reaching down to 5~pc on a side -- far below the 50 pc image pixel size. The number of photon packets launched for both primary and secondary emission ranges from 5 to 25 $\times\,10^9$ and is determined for each galaxy using a heuristic as a function of the number of input particles and the number of pixels in the output images. These discretization settings cause a high level of convergence for spatially integrated quantities calculated from the images. Our tests confirm that variations on integrated luminosities caused by numerical issues are well below one per cent for all wavelengths and sight lines as long as the employed aperture is not exceedingly small.

More importantly, we need to evaluate convergence on a pixel by pixel basis. Following \citet{Kapoor2021}, we calculate the relative error $R$ based on Monte Carlo statistics recorded for each pixel \citep{Camps2020} in a representative set of broadband images at selected viewing angles for all ARTEMIS galaxies. According to \citet{Camps2020}, the pixel value can be considered to be reliable for $R<0.1$, it is questionable in the range $0.1 < R < 0.2$, and it is unreliable for $R > 0.2$. Fig.~\ref{fig:ImageFrames} shows these statistics for the face-on (top half) and edge-on (bottom half) views of a single representative ARTEMIS galaxy in four selected bands from UV to submm wavelengths. 

The synthetic surface brightness maps in the top row of each half of the figure use a logarithmic colour scale; the transition between red and blue marks 1/100 of the maximum surface brightness. Each of the underlying data frames has $1728^2$ $50\times50$~pc pixels, corresponding to the 86~kpc field of view (rounded up to a multiple of 64 pixels). The second row in each half of the figure ($b=1$) shows the corresponding $R$ value for each of these pixels indicating reliable (green), questionable (orange) and unreliable (red) image areas. The third row in each half of the figure ($b=2$) shows the same statistic after $2\times2$ binning into $100\times100$~pc pixels. As expected, this binning results in a substantial increase in reliability at the cost of lower resolution. At the binned $b=2$ level, the reliable area (green) essentially covers all pixels with a value down to 1/100 of the maximum surface brightness (yellow and red). At the original $b=1$ level, one needs to include the questionable area (orange) to achieve a similar coverage.

Inspection of the $R$ values in representative broadbands for the face-on, edge-on and random viewing angles confirms that convergence for the other ARTEMIS galaxies is similar to the results shown in Fig.~\ref{fig:ImageFrames}. We note that the FUV/NUV bands tend to show somewhat poorer statistics because of the relatively lower fluxes and higher extinction involved in that wavelength range.


\begin{figure*}
  \centering
  \includegraphics[width=\textwidth]{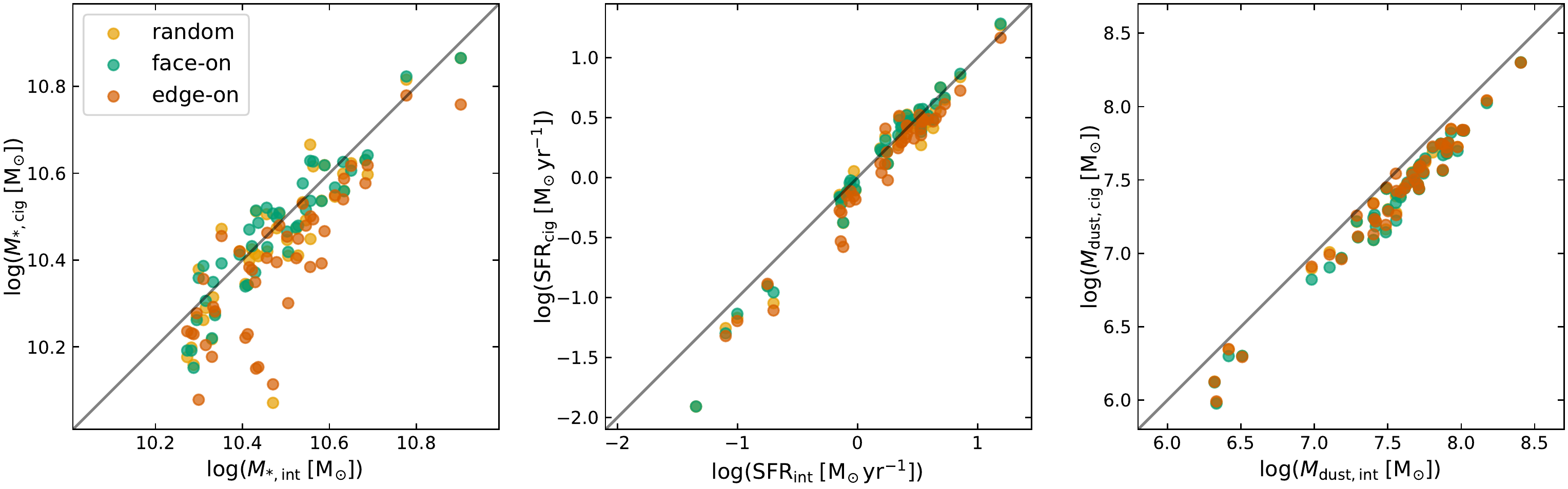}
  \caption{Global physical properties of the ARTEMIS galaxies derived from synthetic observations (vertical axis) for three sight lines (see legend) versus the corresponding intrinsic properties (horizontal axis). From left to right: stellar mass, SFR, and dust mass. The solid line indicates the one-to-one relation. The dustless galaxies G21 and G37 are not shown.}
  \label{fig:DerivedIntrinsic}
\end{figure*}

\begin{figure}
  \centering
  \includegraphics[width=0.93\columnwidth]{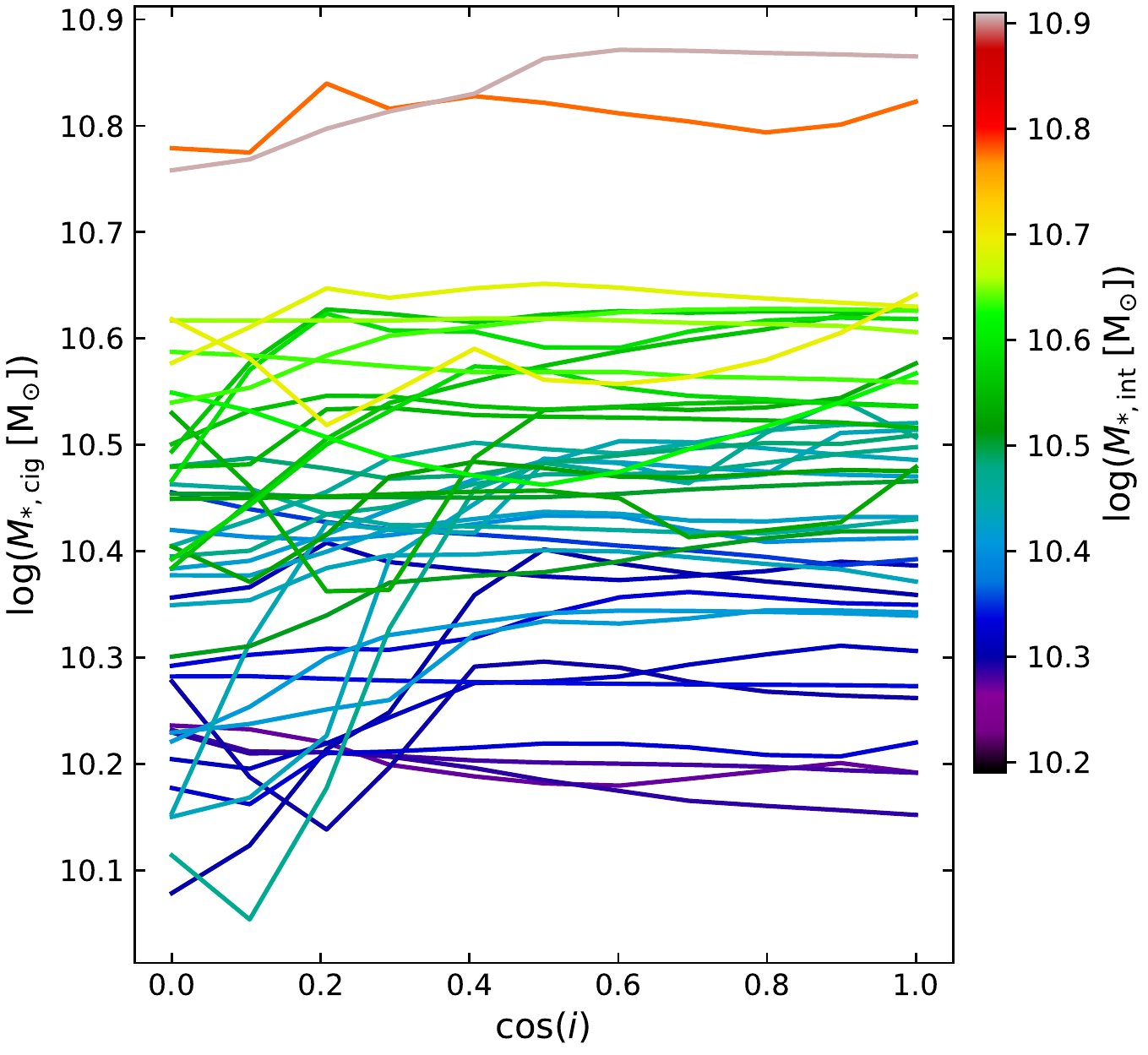}
  \caption{Inferred stellar mass of each ARTEMIS galaxy as a function of the inclination of the synthetic observation from which it has been derived. The curves are colour-coded for the intrinsic stellar mass of the corresponding galaxy, as indicated by the colour bar. The dustless galaxies G21 and G37 are not shown.}
  \label{fig:StellarMassInclination}
\end{figure}

\begin{figure}
  \centering
  \includegraphics[width=0.85\columnwidth]{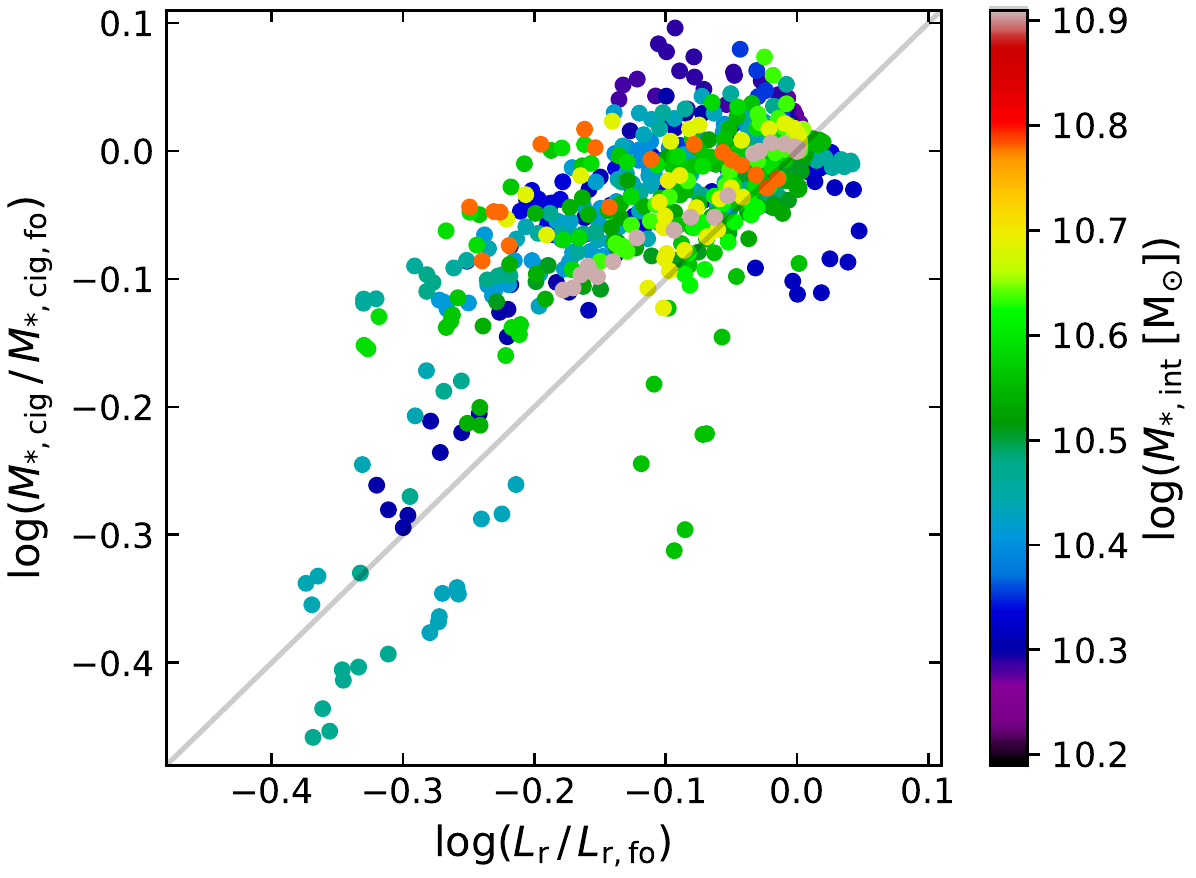}
  \caption{Variation in the inferred stellar mass for various inclinations of the ARTEMIS galaxies (relative to the face-on value), as a function of the corresponding variation in SDSS $r$ luminosity (also relative to the face-on value). The dots are colour-coded for the intrinsic stellar mass of the corresponding galaxy using the same colour scheme as in Fig.~\ref{fig:StellarMassInclination}. The solid line indicates the one-to-one relation. The dustless galaxies G21 and G37 are not shown.}
  \label{fig:StellarMassAttenuation}
\end{figure}

\begin{figure}
  \centering
  \includegraphics[width=0.96\columnwidth]{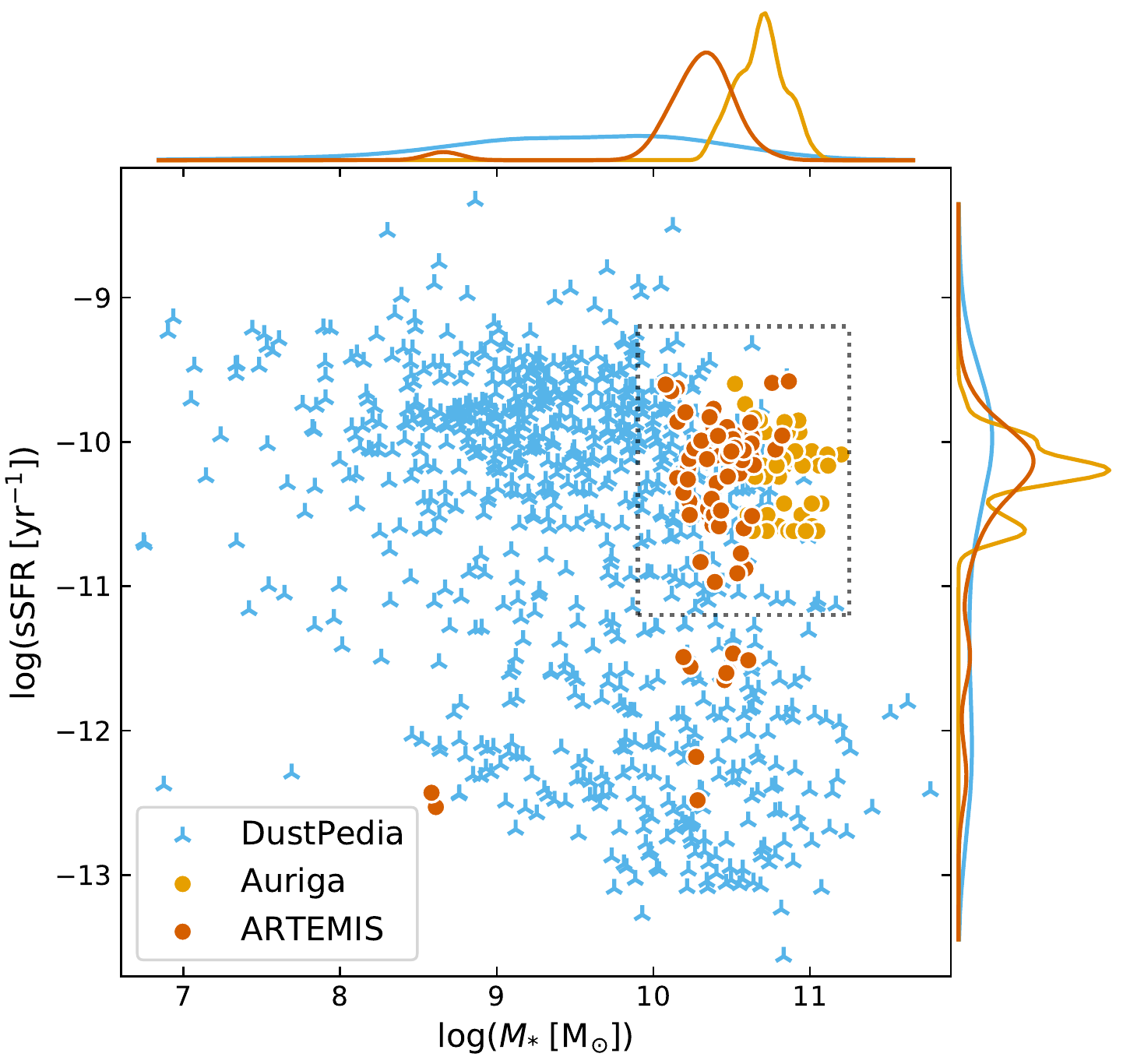}
  \caption{Specific SFR versus stellar mass for the ARTEMIS (red) and Auriga (orange) galaxies over-plotted on the DustPedia observations (blue). The galaxy properties are inferred through SED fitting from synthetic observations for edge-on and face-on sight lines (ARTEMIS and Auriga, using the \sK21/\dT12 recipe) or from actual observations (DustPedia). The non-star-forming ARTEMIS galaxies G21 and G37 are not shown. The dotted rectangle indicates the galaxy selection used for the bottom row of Fig.~\ref{fig:DustScaling}.}
  \label{fig:StarFormationMass}
\end{figure}

\begin{figure*}
  \centering
  \includegraphics[width=0.9\textwidth]{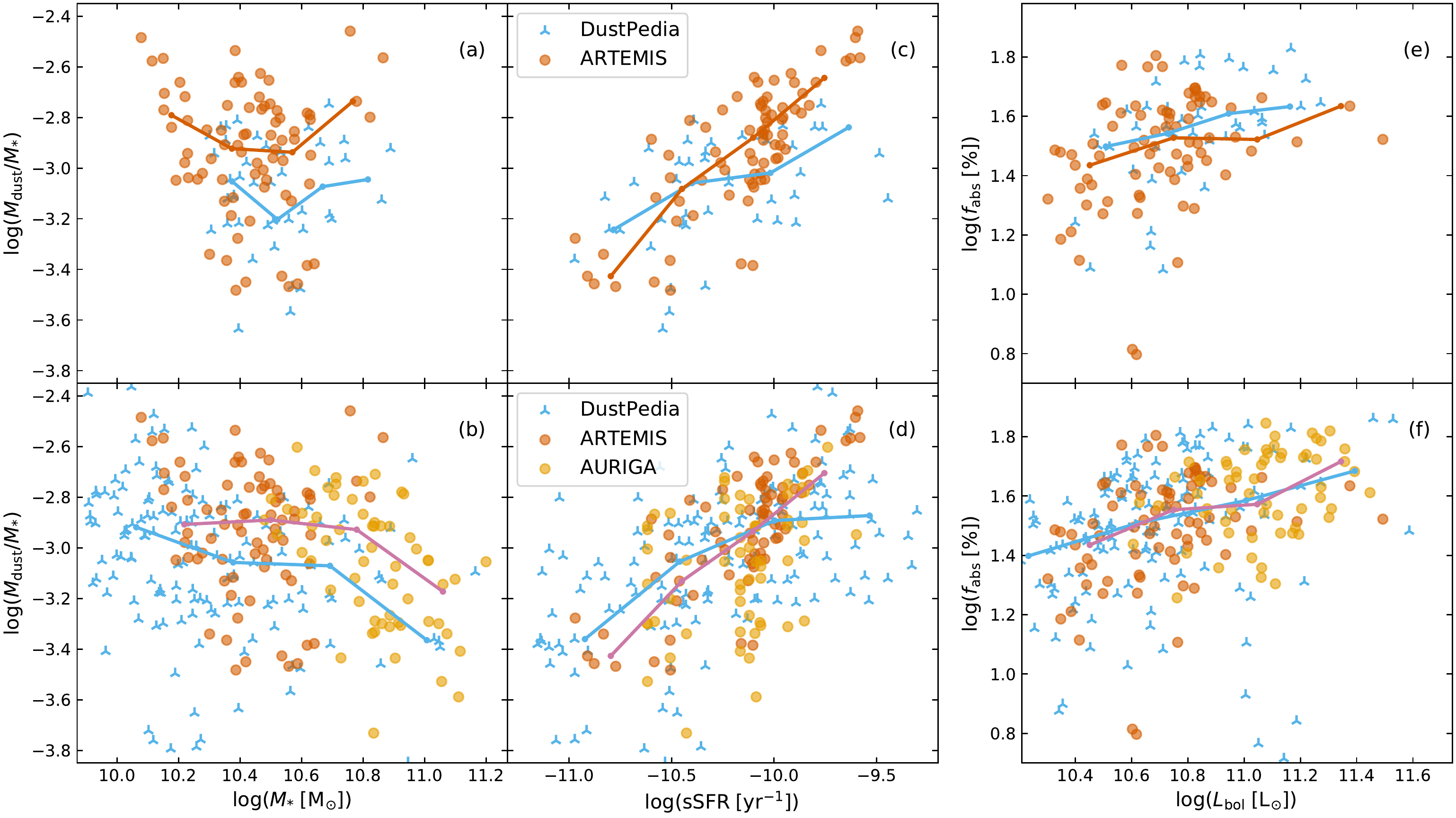}
  \caption{Dust scaling relations for the ARTEMIS (red), AURIGA (orange) and DustPedia (blue) galaxies. The top row compares the matched ARTEMIS and DustPedia samples defined in Sect.~\ref{sec:sampleselection}. The bottom row compares the combined set of ARTEMIS and Auriga galaxies to the DustPedia galaxy subset with sSFR--$M_*$ values inside the dotted rectangle shown in Fig.~\ref{fig:StarFormationMass}. The running median curves trace the combine data sets. The galaxy properties are inferred through SED fitting from synthetic observations for edge-on and face-on sight lines (ARTEMIS and Auriga, using the \sK21/\dT12 recipe) or from actual observations (DustPedia). The first two columns show the specific dust mass versus stellar mass and versus sSFR, respectively. The third column shows the fraction of energy  absorbed by dust as a function of bolometric luminosity.}
  \label{fig:DustScaling}
\end{figure*}


\section{Analysis}
\label{sec:analysis}

\subsection{Global physical properties}
\label{sec:globalprops}

Given the synthetic images described in Sect.~\ref{sec:dataproducts}, we now derive global physical properties for the ARTEMIS galaxies and compare those to observations. As a first step, we calculate global fluxes by integrating the surface brightness maps within the $5\mathrm{R}_{\mathrm{M}50}$ aperture of the galaxy and convert these to luminosities taking into account the assumed model-instrument distance. This yields results similar to those employed during calibration (see Sect.~\ref{sec:calibrationconvergence}), but now we can calculate luminosities for the full complement of 18 sight lines and 50 broadbands. Subsequently, we use the CIGALE SED fitting code \citep{Noll2009, Boquien2019} to estimate stellar mass, SFR, and dust mass from a relevant subset of 25 of these broadband luminosities spanning the UV-submm wavelength range. As described in Sect.~\ref{sec:cigale}, we use the same parameter settings as those used to obtain physical properties of the DustPedia galaxies so that we can compare simulated and observed galaxies on equal footing.

\subsubsection{Inferred versus intrinsic properties}
\label{sec:inferredintrinsic}

As an initial sanity check, Fig.~\ref{fig:DerivedIntrinsic} compares these inferred physical properties to the corresponding intrinsic properties for the ARTEMIS galaxies. The stellar mass (left panel) is estimated accurately within $\pm$0.1~dex from the face-on view but is underestimated by up to 0.4 dex from the edge-on view. We will further investigate the inclination dependence of the stellar mass estimate later in this subsection. The SFR (middle panel) is also estimated well ($\pm$0.25~dex except for one outlier) with a much smaller inclination dependence. For both stellar mass and SFR there is an increasing underestimation for lower stellar mass/SFR values, even for the face-on results. \citet{Kapoor2021} do not see such a trend for the Auriga galaxies (private communication), but this is not really in conflict because the Auriga intrinsic stellar masses are $\gtrsim 10^{10.5}~\Msun$. \citet{Trcka2020} do not find a significant trend with stellar mass in their analysis of the EAGLE galaxies, although their Fig.~4 does show some outliers in the same stellar mass/SFR range. The origin of this discrepancy is unclear. One possible cause is related to the sampling of SF regions. The \sC16 recipe employed for EAGLE splits SF region particles into smaller sub-particles, while the \sK21 recipe employed for both the Auriga and our ARTEMIS results does not. Galaxies with lower stellar mass/SFR necessarily have fewer SF particles, causing a poorer statistical sampling that may lead to systematic effects.

The dust mass (right panel of Fig.~\ref{fig:DerivedIntrinsic}) is systematically underestimated by $0.20\pm0.15$~dex. As expected, there is no significant inclination dependence because thermal dust emission is essentially isotropic. \citet{Kapoor2021} find a similar systematic underestimation of the dust mass for the Auriga galaxies. \citet{Hunt2019} compare methods for fitting SEDs to the most recent photometry \citep{Dale2017} of the nearby star-forming galaxies in the KINGFISH survey \citep{Kennicutt2011}. The methods under study include the SED fitting codes CIGALE and MAGPHYS \citep{daCunha2008} and a method employing a library of SEDs produced from spheroidal GRASIL models \citep{Silva1998} through a RT process. The authors note that GRASIL tends to report dust masses larger than those of CIGALE or MAGPHYS by $\approx$0.3~dex. They subsequently conclude that, because GRASIL is the only method that performs RT in realistic geometries, this may indicate that the other methods are underestimating dust mass.

\citet{Dudzevic2020} perform an analysis similar to ours for the 9431 galaxies at redshift $z>0.25$ and with $\mathrm{SFR} > 10~\Msun\,\mathrm{yr}^{-1}$ in the reference EAGLE simulation \citep{Schaye2015}. The authors use MAGPHYS \citep{daCunha2008} to derive physical properties from synthetic SEDs produced via SKIRT \citep{Camps2018a} and then compare these inferred properties to the intrinsic properties in their supplemental Fig.~A2. Although the galaxies in their analysis are at nonzero redshift and have, on average, a much higher SFR than the galaxies in our study, it is interesting to note a number of differences and similarities. The inferred stellar mass in their analysis is systematically underestimated by $\approx$0.3~dex and the SFR by $\approx$0.1~dex. These systematic deviations might be caused in part by inclination effects (see our Fig.~\ref{fig:DerivedIntrinsic}, left and middle panel; the authors presumably used a random inclination), in addition to different model assumptions in both the radiation transfer (e.g., the dust model) and the SED fitting (e.g., the IMF). On the other hand, the scatter of the inferred properties around the best fitting line is very similar to the scatter in our results, and there is a noticeable trend towards larger stellar mass underestimation for lower-mass galaxies similar to our findings (Fig.~\ref{fig:DerivedIntrinsic}, left panel). The dust mass in their analysis is systematically underestimated by $\approx$0.2~dex, which is very similar to our results (Fig.~\ref{fig:DerivedIntrinsic}, right panel).

We now come back to the inclination dependence of the stellar mass estimate mentioned earlier in this subsection. Fig.~\ref{fig:StellarMassInclination} shows the inferred stellar mass of the ARTEMIS galaxies as a function of inclination. The curves are colour-coded for the intrinsic stellar mass of the corresponding galaxy as indicated by the colour bar. Although for some galaxies the estimated mass seems to dip and rise almost arbitrarily with inclination, most galaxies show a systematic underestimation at high inclinations. Previous studies \citep[e.g.,][]{Malek2018, Trcka2020, Trayford2020} have found that adopting an attenuation curve with a slope that is more shallow than typically assumed observationally can lead to underestimation of the attenuation at optical wavelengths and therefore underestimation of the stellar mass. \citet{Kapoor2021} find a similar trend of underestimation at high inclinations for the Auriga galaxies. They also show that the attenuation curve fitted by the CIGALE code to the Auriga galaxies is significantly more shallow than the attenuation curves observed for the DustPedia galaxies, supporting the above reasoning.

The question remains why this effect is more prominent at high inclinations. Many SED fitting codes, including CIGALE, assume energy balance between the stellar light absorbed by dust and the thermal radiation emitted by dust. However, while the thermal emission at long wavelengths is virtually isotropic, the observed stellar light depends significantly on the dust attenuation experienced along a given sight line, breaking energy balance for that particular line of sight. We can thus intuitively expect the accuracy of SED fitting to depend on the optical attenuation.

To investigate this further, Fig.~\ref{fig:StellarMassAttenuation} shows the variation in the inferred stellar mass for the ARTEMIS galaxies as a function of the corresponding variation in SDSS $r$ luminosity, in both cases relative to the face-on value. Each galaxy is represented by 18 dots, one for each of the simulated sight lines, and these dots are coloured for the galaxy's intrinsic stellar mass as in Fig.~\ref{fig:StellarMassInclination}. The luminosity variation has a zero or negative value in virtually all cases. In other words, as expected, the face-on view used as a reference usually has the highest luminosity, and we can interpret the values on the horizontal axis as a proxy for attenuation at optical wavelengths. Similarly, because the face-on inferred stellar mass correlates well with the intrinsic stellar mass (Fig.~\ref{fig:DerivedIntrinsic}, left panel) we can interpret the values on the vertical axis as a proxy for stellar mass underestimation. With this in mind, Fig.~\ref{fig:StellarMassAttenuation} shows a clear overall correlation between the stellar mass underestimation by the SED fitting algorithm and the attenuation for a given sight line. Many individual galaxies show the same trend, with multiple dots forming an approximate line, often with a similar slope as the overall trend. On the other hand, there is a significant amount of scatter on the relation ($\approx 0.3$ dex). This is not surprising, given that the observed attenuation will depend substantially on the precise star-dust geometry, especially in near-edge-on configurations.

\subsubsection{Dust scaling relations}

We now compare simulated and observed data sets using galaxy properties inferred through SED fitting of synthetic fluxes (ARTEMIS and Auriga) or observed fluxes (DustPedia). Fig.~\ref{fig:StarFormationMass} presents our three data sets in the sSFR--$M_*$ plane. The figure shows all galaxies with sSFR and $M_*$ above the lower axis limits. This excludes just a few low-mass DustPedia galaxies, two non-star-forming ARTEMIS galaxies, and no Auriga galaxies. 

The Auriga zoom simulations are selected using a halo mass cutoff of $1\times10^{12} < M_{200}/\Msun < 2\times10^{12}$ in addition to a requirement of relative environmental isolation (see Sect~\ref{sec:auriga}). The resulting galaxies consequently occupy a fairly limited region in the upper right corner of the sSFR--$M_*$ plane (Fig.~\ref{fig:StarFormationMass}). The ARTEMIS zoom simulations use a more relaxed halo mass cutoff of $8\times10^{11} < M_{200}/\Msun < 2\times10^{12}$ without any environmental criteria (see Sect~\ref{sec:artemis}). The resulting ARTEMIS data set includes galaxies in a lower stellar mass range. Considering only active galaxies with $\mathrm{sSFR}>10^{-11}~\mathrm{yr}^{-1}$, the low end of the stellar mass range is decreased from $\approx 3\times10^{10}~\Msun$ to $\approx 1\times10^{10}~\Msun$. It is evident from Fig.~\ref{fig:StarFormationMass} that the ARTEMIS data set supplements the Auriga data set in this manner. Although there are also a few passive ARTEMIS galaxies with $\mathrm{sSFR}<10^{-11}~\mathrm{yr}^{-1}$, this region of the sSFR--$M_*$ plane remains largely under-sampled. 

Fig.~\ref{fig:DustScaling} shows dust scaling relations for our data sets. The top row compares the matched ARTEMIS and DustPedia samples defined in Sect.~\ref{sec:sampleselection}. The bottom row compares the combined set of ARTEMIS and Auriga galaxies to a DustPedia galaxy subset defined by the sSFR and $M_*$ limits indicated by the dotted rectangle in Fig.~\ref{fig:StarFormationMass}. These limits have been chosen to enclose the DustPedia galaxies in the sample matching ARTEMIS (defined in Sect.~\ref{sec:sampleselection}) and those in the sample matching Auriga \citep[defined by][DPD45]{Kapoor2021}. These panels again illustrate how the ARTEMIS and Auriga data sets supplement each other, although we do need to keep in mind that they originate from simulations with different assumptions and subgrid physics. Generally speaking, the synthetic galaxy scaling relations correspond to the observations fairly well. We now discuss each column in turn.

The first column of Fig.~\ref{fig:DustScaling} shows specific dust mass versus stellar mass. We recall from Sect.~\ref{sec:sampleselection} that the DustPedia sample in panel (a) has been selected using the intrinsic ARTEMIS stellar mass range. As discussed in Sect.~\ref{sec:inferredintrinsic} and illustrated in Fig.~\ref{fig:DerivedIntrinsic}, the CIGALE SED fitting procedure underestimates the intrinsic stellar mass by $0.20\pm0.15$~dex. This causes a corresponding shift to lower masses of the ARTEMIS galaxies compared to the DustPedia sample in panel (a) of Fig.~\ref{fig:DustScaling}. Furthermore, for a given stellar mass, the ARTEMIS specific dust mass is higher than that observed for DustPedia by $\approx 0.3$~dex. The expected decrease of specific dust mass with increasing stellar mass is not seen in this panel, possibly because of the narrow stellar mass range and the restricted sSFR range (limiting the effect of an increasing passive fraction with stellar mass). The downward trend in the relation is recovered when combining the ARTEMIS and Auriga galaxies as shown in panel (b), although the specific dust mass for the synthetic galaxies generally remains too high. \citet{Kapoor2021} argue that this discrepancy might be caused by a high gas content of the simulated galaxies compared to observations rather than by issues in the post-processing procedure. 

The second column of Fig.~\ref{fig:DustScaling} shows specific dust mass versus sSFR. The observed DustPedia relation is reproduced well by both the ARTEMIS (panel c and d) and the Auriga galaxies (panel d), although there is a slight discrepancy in the slope. Compared to observations, the synthetic galaxies show a slightly more rapid increase in specific dust mass with increasing sSFR. In any case, the panels confirm a clear correlation between (specific) dust mass and (specific) SFR.

The third column of Fig.~\ref{fig:DustScaling} shows the fraction of energy absorbed by dust, defined as $f_\mathrm{abs} = L_\mathrm{dust} / L_\mathrm{bol}$, versus bolometric luminosity. The observed DustPedia relation and scatter are reproduced excellently by both the ARTEMIS (panel e and f) and the Auriga galaxies (panel f). The Auriga galaxies are, on average, more luminous than the ARTEMIS galaxies, so that the two simulated data sets occupy largely distinct regions in the $f_\mathrm{abs}$--$L_\mathrm{bol}$ plane (panel f). Still, the slope of the $f_\mathrm{abs}$--$L_\mathrm{bol}$ relation for the combined simulated data set very closely follows the observed slope across the full luminosity range.


\begin{figure*}
  \centering
  \includegraphics[width=\textwidth]{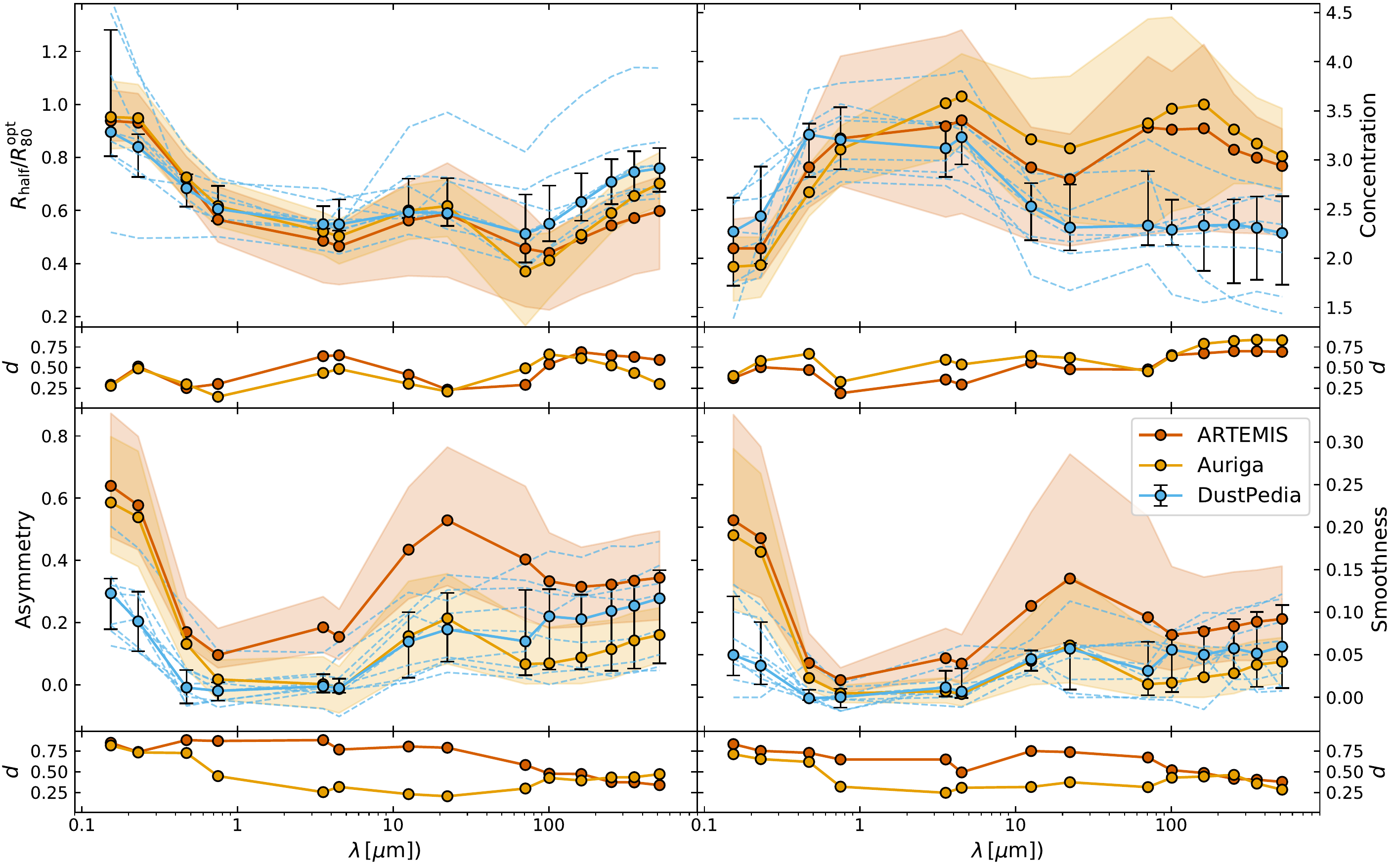}
  \caption{Morphological indices (half light radius, concentration, asymmetry, and smoothness; see Sect.~\ref{sec:statmorph} for definitions) as a function of wavelength for three sets of disc galaxies. The 9 DustPedia galaxies (blue) correspond to those studied by \citet{Baes2020}. The Auriga data set (orange) corresponds to the 14 galaxies studied by \citet{Kapoor2021}, and the ARTEMIS data represent the 11 ARTEMIS galaxies with a disc-to-total stellar mass ratio of $\mathrm{D/T}>0.45$, using the \sK21/\dT12 recipe for 5 inclinations ranging from face-on to $i=73^\circ$ in both cases. The circular markers represent the median values for each set; the error bars (DustPedia) or shaded areas (synthetic data sets) indicate the $\pm1\sigma$ interval. The dashed lines show individual DustPedia galaxies. The small sub-panels under each panel show the K--S test distance $d$ as a function of wavelength, quantifying the dissimilarity between synthetic and observed data sets, i.e. Auriga versus DustPedia (orange) and i.e. ARTEMIS versus DustPedia (red).}
  \label{fig:MorphologyIndices}
\end{figure*}


\subsection{Morphology on resolved scales}
\label{sec:morphology}

In this section we investigate selected non-parametric morphological parameters of the ARTEMIS disc galaxies as a function of wavelength, comparing the results to DustPedia observations and to the simulated Auriga galaxies. We use the StatMorph package \citep{RodriguezGomez2019} to derive the CAS indices (concentration, asymmetry, smoothness) and the normalised elliptical half light radius ($R_\mathrm{half}/R_{80}^\mathrm{opt}$) from the ARTEMIS images in 14 broadbands ranging from UV to submm wavelengths. Sect.~\ref{sec:statmorph} offers some background on the StatMorph code and a concise definition of the morphological indices used here.

We use the same wavelength bands as those employed by \citet{Baes2020} to study the morphology of 9 well-resolved spiral galaxies in the DustPedia database and by \citet{Kapoor2021} to study a set of 14 disc galaxies from the Auriga simulations. Following \citet{Kapoor2021}, we select ARTEMIS galaxies with a disc-to-total stellar mass ratio of $\mathrm{D/T}>0.45$, using the intrinsic stellar mass values listed by \citet[][Table~1]{Font2020}. This yields a set of 11 galaxies. Just as was done for the Auriga galaxies, we employ synthetic images for 5 inclinations ranging from face-on to $i=73^\circ$ for each galaxy, leading to $5\times11$ data points for each wavelength. For both synthetic data sets, the images have been produced using the \sK21/\dT12 recipe (see Sect.~\ref{sec:combinedrecipes}). The ARTEMIS image field of view encloses the galaxy's full extraction aperture (see Sect.~\ref{sec:aperture} which always includes the stellar surface density-based aperture used by \citet{Kapoor2021} for Auriga.

Fig.~\ref{fig:MorphologyIndices} shows the four morphological indices under study as a function of wavelength for these three sets of disc galaxies. The circular markers represent the median values for each set; the error bars (DustPedia) or shaded areas (ARTEMIS, Auriga) indicate the $\pm1\sigma$ interval. The dashed lines show individual DustPedia galaxies. The small sub-panels under each panel show a metric $d$ quantifying the `distance' between synthetic and observed data sets as a function of wavelength. This metric is calculated using the 1D two-sample Kolmogorov-Smirnov test \citep[K--S test,][]{Kolmogorov1933,Smirnov1948}. 

The observed normalised half light radius $R_\mathrm{half}/R_{80}^\mathrm{opt}$ (upper left panel of Fig.~\ref{fig:MorphologyIndices}) shows a characteristic trend as a function of wavelength, with large values in the FUV and a gradual decrease over the optical regime to the NIR, followed by a small increase in the MIR and another dip before a final increase towards FIR and submm wavelengths. Both the Auriga and the ARTEMIS simulations reproduce the observed DustPedia trend well, although the median ARTEMIS radii are consistently smaller than the corresponding Auriga values and, for wavelengths longer than optical, also smaller than the DustPedia values. The difference is most notable in the FIR wavelength range corresponding to dust emission, implying that ARTEMIS dust discs are smaller relative to their stellar discs than those in the Auriga simulations and the DustPedia galaxies.

The observed concentration index (upper right panel of Fig.~\ref{fig:MorphologyIndices}) shows a characteristic trend with high concentration for NIR and MIR wavelengths, tracing older stellar populations, and substantially lower concentration for shorter and longer wavelengths, tracing younger stellar populations and dust. The ARTEMIS and Auriga simulations reproduce the observed DustPedia trend fairly well for wavelengths shorter than $\approx 10~\mu\mathrm{m}$. For longer wavelengths, both simulations show substantially higher concentration than observed. Despite this discrepancy, the ARTEMIS concentration values are consistently closer to the DustPedia values than the Auriga results, particularly in wavelength regimes that trace dust either through extinction (UV and optical) or emission (FIR and submm). Because the same dust allocation scheme has been used for post-processing ARTEMIS and Auriga galaxies, and given the constant dust-to-metal ratio in this recipe, this seems to indicate that the ARTEMIS simulations include a slightly better prediction of the metal distribution in the galaxy, although still falling short of observations (at longer wavelengths, the two simulations are much closer to each other than to the data). Because there are many differences between the subgrid recipes of the two simulations, it is hard to pin down the precise cause. Most likely, differences in the stellar and AGN feedback mechanisms play a significant role, as these processes affect the metal distribution in various ways \citep[e.g., for the effect of AGN feedback on the resolved distribution of metals in the EAGLE simulations, see][]{Trayford2019}.

In the UV and optical wavelength range, the asymmetry and smoothness indices (bottom row of Fig.~\ref{fig:MorphologyIndices}) for both ARTEMIS and Auriga are very similar to each other and both are substantially higher than observed for DustPedia. These high values are probably caused by the SF regions, which are prominent at those wavelengths (see Fig.~\ref{fig:ImageFrames} for two examples). As noted by \citet{Kapoor2021}, these index values may improve in case the SF regions would be resampled during post-processing as in recipe \sC16 (see Sect.~\ref{sec:sfrrecipes}). At longer wavelengths, starting from the NIR, the ARTEMIS galaxies continue to show asymmetry and smoothness values well above the DustPedia reference values, while the Auriga galaxies are very close to observations. Notably, for wavelengths longer than $\approx 100~\mu\mathrm{m}$, the ARTEMIS galaxies continue to over-predict while the Auriga galaxies under-predict by roughly the same amount. This discrepancy might be related to differences in the merger histories for the two simulated galaxy sets. ARTEMIS includes histories with more recent mergers, whereas Auriga has a criterion for isolation.

All in all, the morphology of our ARTEMIS galaxy sample follows the same overall trends as a similar DustPedia galaxy sample, with some notable discrepancies. Using the same post-processing recipe, the ARTEMIS simulations provide better predictions for the concentration index than the Auriga simulations but falls short for the other indices under study, at least in some wavelength regimes.


\section{Summary and conclusions}
\label{sec:conclusions}

In this work, we produce and publish multi-wavelength, spatially resolved synthetic observations for the 45 simulated galaxies of the ARTEMIS project \citep{Font2020, Font2021}. These galaxies were selected to have a Milky Way-like halo mass and were re-evolved to redshift zero at high resolution including full (subgrid) baryonic physics. We extract stellar and gas properties from the present-day galaxy snapshots with the purpose of calculating synthetic observables with our RT code SKIRT \citep{Camps2015a, Camps2020}. We assign emission spectra and dust characteristics following a combination of previously developed prescriptions \citep{Camps2016, Trayford2017, Kapoor2021}. Stellar populations are modelled through the \citet{Bruzual2003} template library. We include a subgrid treatment of SF regions using the MAPPINGS III template library \citep{Groves2008} to help capture the dust RT processes in their dense and clumpy cores. We allocate diffuse interstellar dust to the galaxy's cold gas using a fixed dust-to-metal ratio, $f_\mathrm{dust}$, which is treated as a free parameter. The dust properties are taken from the THEMIS dust model \citep{Jones2017}. We explore variations of the recipe with or without re-sampling of the SF regions and with a more concentrated or more extended dust allocation scheme.

We calibrate the value of $f_\mathrm{dust}$ for each recipe variation by comparison with observed galaxies in the DustPedia database \citep{Davies2017}. We construct mutually matched samples including the 38 star-forming ARTEMIS galaxies on the one hand and 42 DustPedia galaxies in the same stellar mass and SFR range on the other hand. We then compare these samples through luminosity scaling relations in wavelength bands from UV to submm (Fig.~\ref{fig:scalingrelations}). The resulting optimal $f_\mathrm{dust}$ values for each of our dust allocation recipes are listed in Table~\ref{tab:fdust}. We furthermore conclude that the \sK21/\dT12 recipe optimally reproduces the observed luminosity scaling relations, confirming the findings by \citet{Kapoor2021} for the simulated Auriga galaxies, albeit with a different value for $f_\mathrm{dust}$ (see Table~\ref{tab:fdust}).

Even with the optimal recipe, dust extinction is significantly underestimated at FUV/UV wavelengths and representative dust temperatures are lower than those observed. We attribute these discrepancies to limitations in the treatment of SF regions and their immediate environment, similar to the findings in previous studies \citep[e.g.,][]{Camps2016, Trcka2020, Kapoor2021}. These symptoms therefore seem to be a characteristic of all state-of-the-art cosmological simulation UV-submm post-processing efforts using similar SF region recipes. Resolving these issues will likely require improvements both in the modelling of the cold interstellar medium in hydrodynamical simulations of galaxy evolution and in the subgrid treatment of SF regions in the RT post-processing procedure. Concerning the latter, a crucial step is to develop an enhanced SF region model that is designed specifically for incorporation in RT simulations. Ideally, key characteristics of the model such as the SSP or dust grain properties should be configurable, and the parameters of the resulting SED template library should be easily derivable from the particle properties in the hydrodynamical simulation. More fundamentally, the model should, on average, allow more UV radiation to escape into the diffuse ISM without adversely affecting the shape of the SED at other wavelengths.

Using the optimal \sK21/\dT12 recipe, we produce images of all ARTEMIS galaxies at 50~pc resolution for 50 commonly-used broadband filters from UV to submm wavelengths and for 18 different viewing angles. We spatially integrate these images to obtain global fluxes and use the SED fitting code CIGALE \citep{Noll2009, Boquien2019} to derive physical galaxy properties. The inferred properties recover the known intrinsic values fairly well, except that stellar mass is often significantly underestimated for near-edge-on configurations (Figs.~\ref{fig:DerivedIntrinsic} and \ref{fig:StellarMassInclination}). We argue that this discrepancy is related to the stronger optical dust attenuation at high inclinations (Fig.~\ref{fig:StellarMassAttenuation}), which disturbs the energy balance for those sight lines, in turn confusing the SED fitting algorithm.

We use selected dust scaling relations (Fig.~\ref{fig:DustScaling}) to compare the inferred ARTEMIS galaxy properties to similarly derived properties of the DustPedia galaxies and of the simulated Auriga galaxies \citep{Kapoor2021}. We find that the ARTEMIS galaxies tend to contain more dust than comparable DustPedia galaxies, but otherwise follow the observed dust scaling relations very well. The Auriga galaxies follow the same relations but often occupy an adjacent region of the parameter space.

We subsequently use the high-resolution images at multiple wavelengths to perform a basic morphological study of the 11 ARTEMIS galaxies with a disc-to-total stellar mass ratio of $\mathrm{D/T}>0.45$. We use the StatMorph package \citep{RodriguezGomez2019} to calculate four non-parametric morphological properties as a function of wavelength. We compare these results (Fig.~\ref{fig:MorphologyIndices}) to similarly derived values for 9 well-resolved spiral galaxies in the DustPedia database and 14 simulated Auriga galaxies. We find that the ARTEMIS galaxies largely reproduce the observed trends as a function of wavelength, except that they appear to be more clumpy and less symmetrical than observed. We also highlight some differences between the ARTEMIS and Auriga data sets.

\citet{Kapoor2021} cite various types of studies of the dust-related properties of simulated Milky Way-like galaxies at redshift zero that are enabled by the availability of dust-aware high resolution images of these galaxies at multiple wavelengths. Their examples include spatially resolved SED fitting, local energy balance studies, spatially resolved dust scaling relations, dust mass maps, and the contribution to dust heating by different stellar components in various regions of a galaxy.

We similarly invite any interested party to use our ARTEMIS results in such studies (see Sect.~\ref{sec:dataproducts}). In fact, the Auriga and ARTEMIS galaxies occupy adjacent regions in the sSFR versus stellar mass plane (Fig.~\ref{fig:StarFormationMass}) and hence also in the dust scaling relations (bottom row of Fig.~\ref{fig:DustScaling}). This means that the data products resulting from this work are supplemental to those produced by \citet{Kapoor2021} for Auriga. Both data sets are publicly available, and because the same post-processing recipe has been used in both cases, they can be combined to achieve a wider coverage of galaxy types and/or to increase statistical significance.


\section*{Acknowledgements}

We thank the anonymous referee for their helpful comments that improved the paper.
AUK and AT acknowledge the financial support of the Flemish Fund for Scientific Research (FWO-Vlaanderen), research projects G039216N and G030319N.
This project has received funding from the European Research Council (ERC) under the European Union’s Horizon 2020 research and innovation programme (grant agreement No 769130).
This work used the DiRAC@Durham facility managed by the Institute for Computational Cosmology on behalf of the STFC DiRAC HPC Facility (www.dirac.ac.uk). The equipment was funded by BEIS capital funding via STFC capital grants ST/K00042X/1, ST/P002293/1, ST/R002371/1 and ST/S002502/1, Durham University and STFC operations grant ST/R000832/1. DiRAC is part of the National e-Infrastructure.

\section*{Data Availability}

Sect.~\ref{sec:dataproducts} describes the public data products provided as a result of this work, available for download at \url{https://www.astro.ljmu.ac.uk/Artemis}.


\clearpage
\bibliographystyle{mnras}
\bibliography{artemis}


\clearpage
\appendix

\section{Comparing dust models}
\label{sec:dustmodels}

\begin{figure}
  \centering
  \includegraphics[width=0.90\columnwidth]{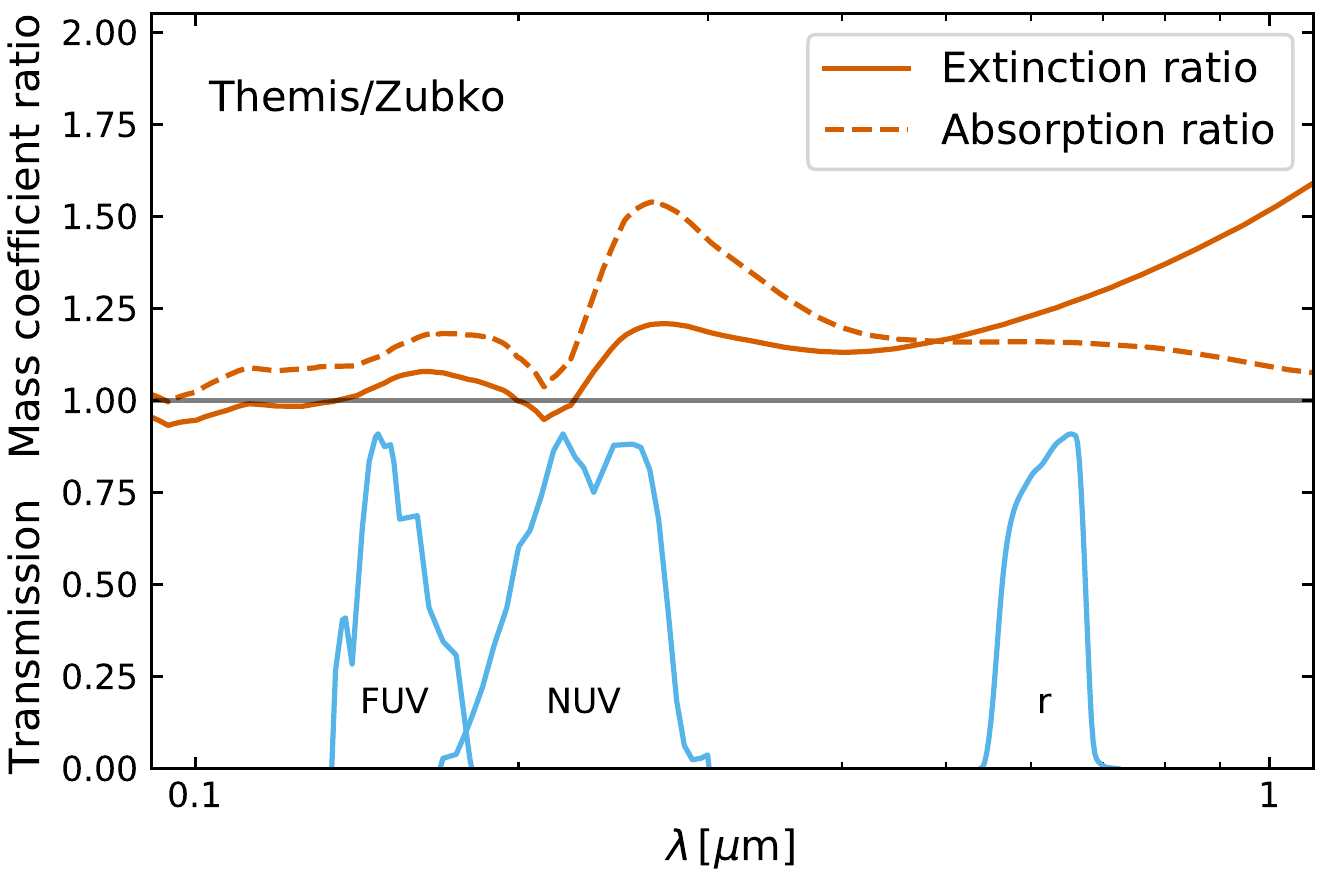}
  \caption{Ratio of the total extinction (solid red) and absorption (dashed red) mass coefficients for the THEMIS \citep{Jones2017} dust model over those for the \citet{Zubko2004} dust model in the UV-optical wavelength range. The bottom panel shows the transmission curves (blue) for the broadbands used in the scaling relations of Fig.~\ref{fig:scalingrelationsdustmodel}.}
  \label{fig:extinction}
\end{figure}

\begin{figure}
  \centering
  \includegraphics[width=0.85\columnwidth]{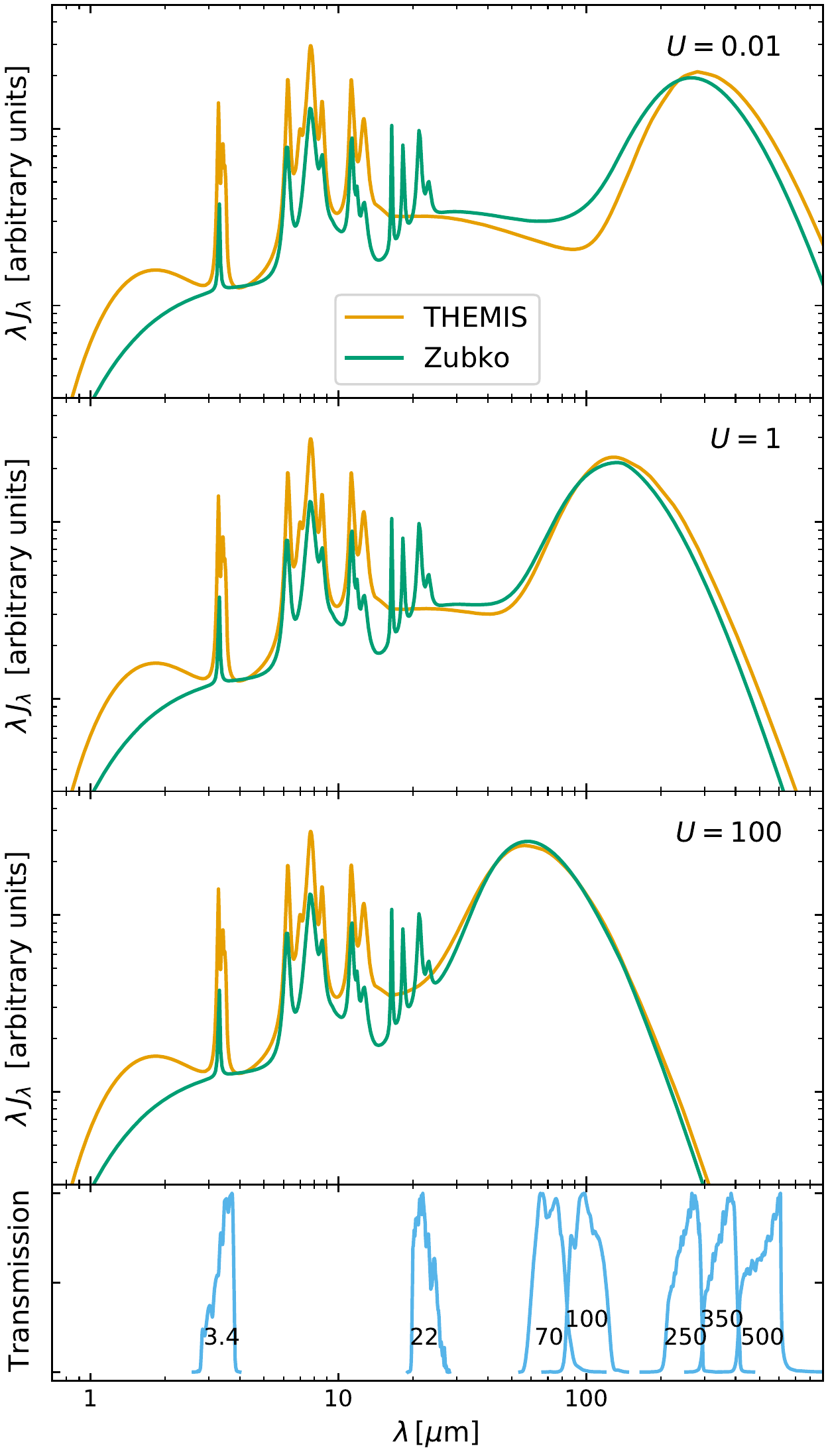}
  \caption{Emissivity of a dust grain population with properties of the THEMIS \citep[][orange]{Jones2017} and \citet[][green]{Zubko2004} dust models in response to a typical interstellar radiation field \citep{Mathis1983} with various strengths $U=0.01, 1, 100$. Details on the emission calculation, including non-equilibrium heating of smaller dust grains, are provided by \citet{Camps2015b} and references therein. The bottom panel shows the transmission curves (blue) for the broadbands used in the scaling relations of Fig.~\ref{fig:scalingrelationsdustmodel}.}
  \label{fig:emissivity}
\end{figure}

\begin{figure*}
  \centering
  \includegraphics[width=\textwidth]{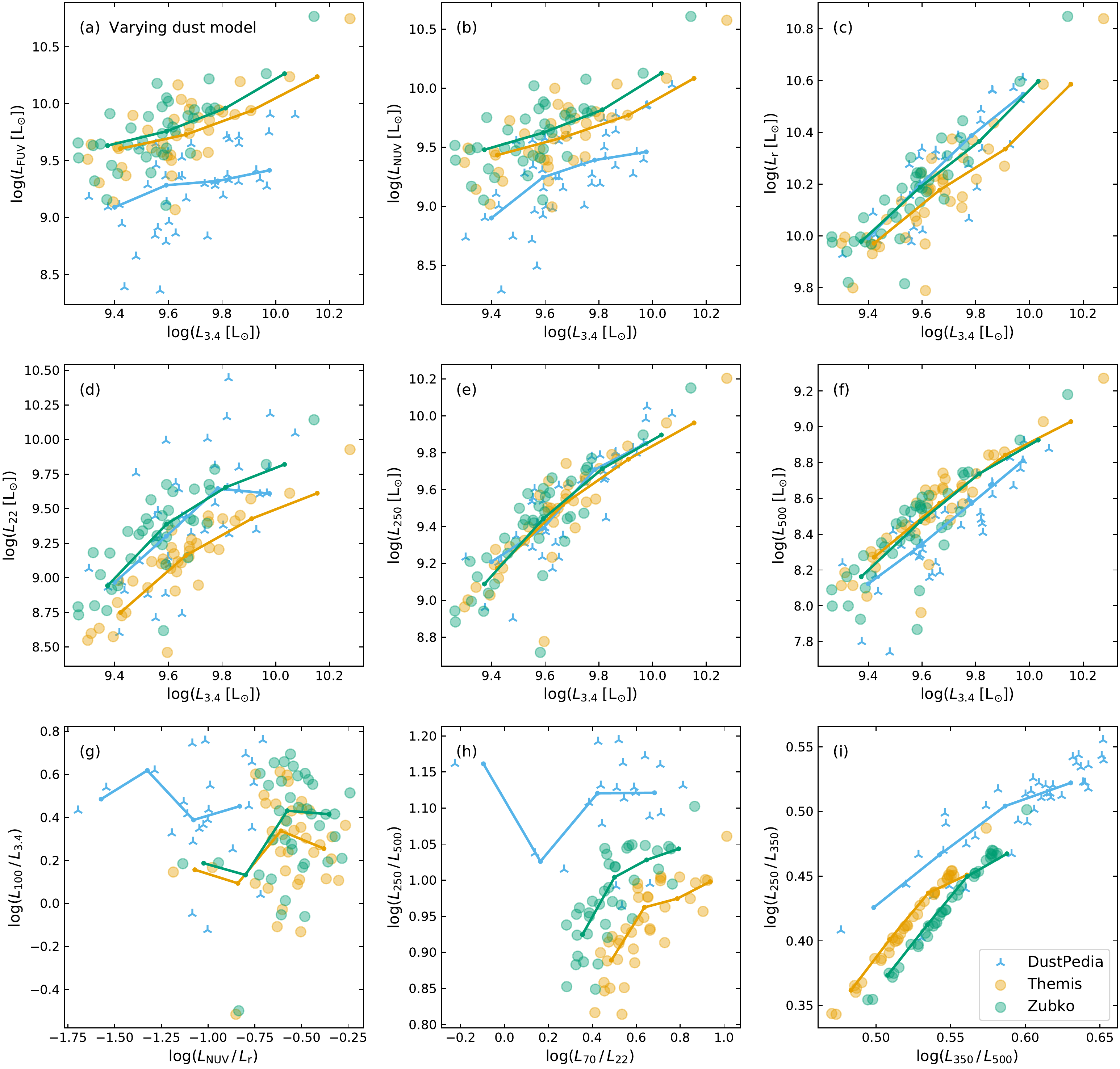}
  \caption{The same scaling relations as in Fig.~\ref{fig:scalingrelations}, now showing ARTEMIS luminosities calculated for a random viewing angle using the \sC16/\dC16 recipe at optimal dust fraction $f_\mathrm{dust}$ (see Table~\ref{tab:fdust}) with THEMIS \citep[][orange]{Jones2017} and \citet[][green]{Zubko2004} dust models in comparison with observed DustPedia luminosities (blue).}
  \label{fig:scalingrelationsdustmodel}
\end{figure*}

As stated in Sect.~\ref{sec:commonrecipes}, we use a more recent dust model for representing the diffuse dust in our RT post-processing procedure than the dust model employed by \citet{Camps2016,Trayford2017} and \citet{Camps2018a} for producing synthetic observables for the EAGLE galaxies. In this appendix we study the effects of this new dust model on the scaling relations shown in Fig.~\ref{fig:scalingrelations}, which we use to calibrate and compare variations of the recipes in Sect.~\ref{sec:calibration}. 

The earlier EAGLE work used the dust model presented by \citet{Zubko2004}, called Zubko in this appendix. This model includes a mixture of non-composite graphite and silicate grains and neutral and ionised polycyclic aromatic hydrocarbon (PAH) molecules, designed so that the global dust properties reproduce the extinction, emission and abundance constraints of the Milky Way. The optical and calorimetric properties follow the prescriptions of \citet{Draine2001} and \citet{Li2001}. In this work, we use the THEMIS dust model described by \citet{Jones2017} and references therein. This model was developed in the context of the DustPedia project to explain the dust extinction and emission in the diffuse interstellar medium, and to self-consistently include the effects of dust evolution in the transition to denser regions. It includes a mixture of amorphous hydrocarbons and amorphous silicates. For the latter, it is assumed that half of the mass is amorphous enstatite and the remaining half is amorphous forsterite.

Fig.~\ref{fig:extinction} shows the ratio of the extinction and absorption mass coefficients for the THEMIS dust model over those for the Zubko dust model in the UV-optical wavelength range. Fig.~\ref{fig:emissivity} compares the emissivity of a THEMIS and Zubko dust grain population of the same mass in response to an interstellar radiation field of varying strength. In each figure, the bottom panel shows the transmission curves for the broadbands used in our scaling relations. From the first figure we conclude that a THEMIS dust grain population exhibits 10 to 25 per cent more extinction than a Zubko grain population in the UV and optical bands under consideration. In the second figure, it is immediately obvious that the aromatic features, on average, are more luminous for the THEMIS dust than for the Zubko dust. This significantly boosts the $3.4~\mu\mathrm{m}$ band luminosity, while the $22~\mu\mathrm{m}$ band luminosity is somewhat tempered. Furthermore, the dust continuum emission peak for weaker input fields shifts slightly to longer wavelengths for the THEMIS model, corresponding to lower representative dust temperatures. Also, the slopes on both sides of the continuum emission peak differ between the dust models. 

With this information we can interpret Fig.~\ref{fig:scalingrelationsdustmodel}, which shows the scaling relations first presented in Fig.~\ref{fig:scalingrelations} for each of the two dust models. The top and middle row panels all have the $3.4~\mu\mathrm{m}$ band luminosity on the horizontal axis. The THEMIS data points in these panels are therefore shifted to the right in accordance with the extra aromatic feature emission modelled in this band (by up to 0.15 dex for the most dust-luminous galaxy). The top row panels have a UV or optical band on the vertical axis, causing the THEMIS data points to shift downward reflecting the increased dust attenuation in that model. The combined result in these panels is a diagonal shift more or less orthogonal to the scaling relation. As discussed in Sect.~\ref{sec:scalingrelations}, our fiducial recipe underestimates attenuation at UV wavelengths. We see here that the THEMIS dust mix helps decreasing this discrepancy. Unfortunately, because its extinction coefficient increases more distinctly for optical wavelengths than for UV wavelengths (see Fig.~\ref{fig:extinction}), the THEMIS model instead overestimates the attenuation in the optical regime.

The THEMIS $22~\mu\mathrm{m}$ data points in panel (d) are also shifted downward, resulting in a diagonal shift similar to that in the UV and optical regimes, but now caused by the decreased aromatic feature emission in this band. The $250~\mu\mathrm{m}$ and $500~\mu\mathrm{m}$ bands are on the downward slope of the dust continuum peak and thus both see enhanced emission for the THEMIS dust model (see Fig.~\ref{fig:emissivity}). As a result, the THEMIS data points in panels (e) and (f) shift essentially along the observed scaling relation. The shifts in the colour-colour relations on the bottom row can be similarly interpreted. Notably, panel (i) clearly shows a lower representative dust temperature for the THEMIS dust mix \citep[the dust temperature rises diagonally to the upper right in this panel; see, e.g., Fig.~11 of][]{Camps2016}.

In summary, compared to the Zubko dust model, and for an otherwise fixed recipe, the THEMIS dust model reduces the discrepancies between our simulations and the DustPedia observations in some wavelength regimes but introduces extra tension in other regimes.

\section{Calibrating dust fractions}
\label{sec:dustfractions}

\begin{figure*}
  \centering
  \includegraphics[width=\textwidth]{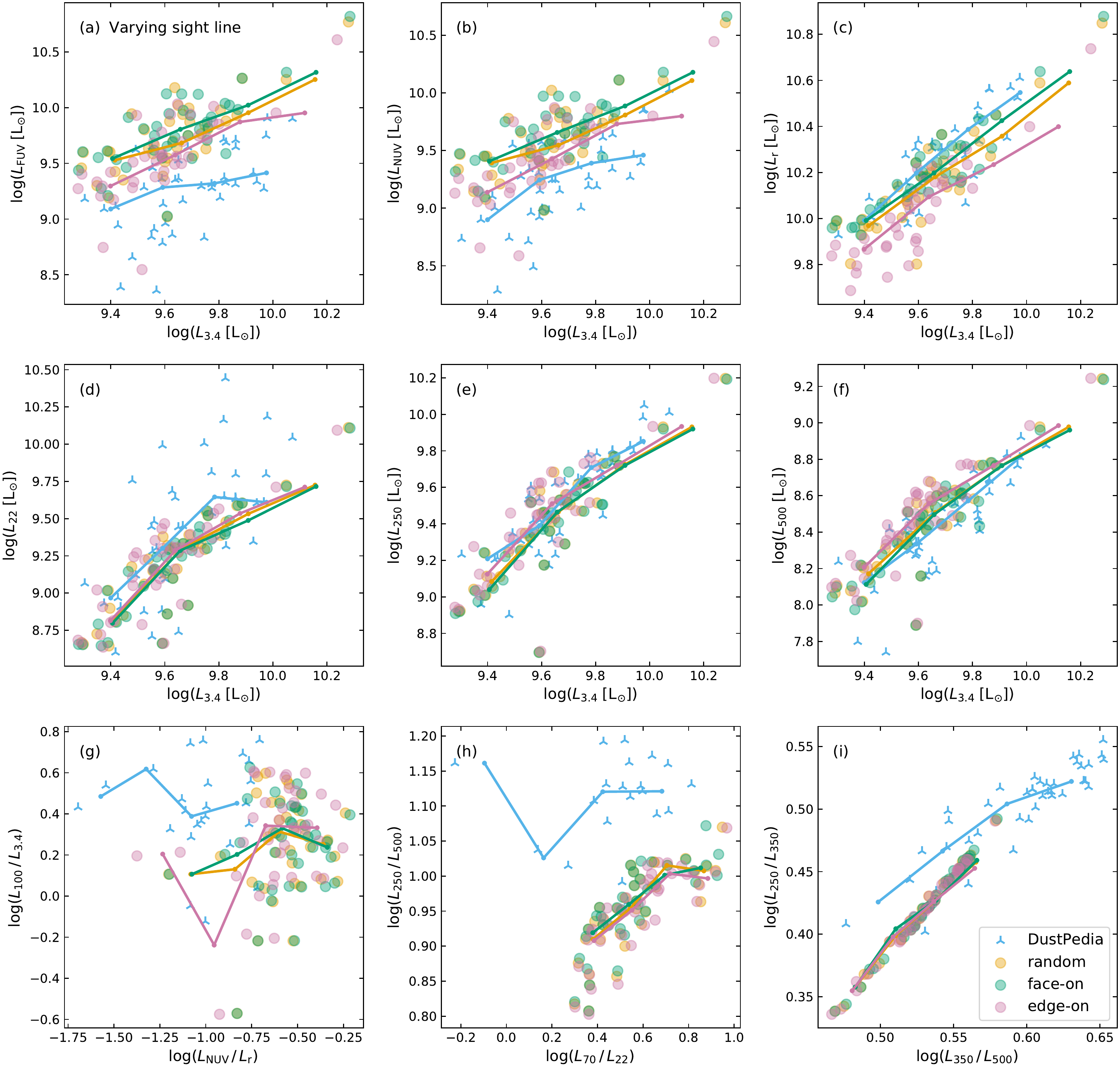}
  \caption{The same scaling relations as in Fig.~\ref{fig:scalingrelations}, now showing ARTEMIS luminosities calculated for three different sight lines (random -- orange, face-on -- green, edge-on -- purple) using the \sK21/\dT12 recipe at optimal dust fraction $f_\mathrm{dust}$ (see Table~\ref{tab:fdust}) in comparison with observed DustPedia luminosities (blue).}
  \label{fig:scalingrelationssightline}
\end{figure*}

\begin{figure*}
  \centering
  \includegraphics[width=\textwidth]{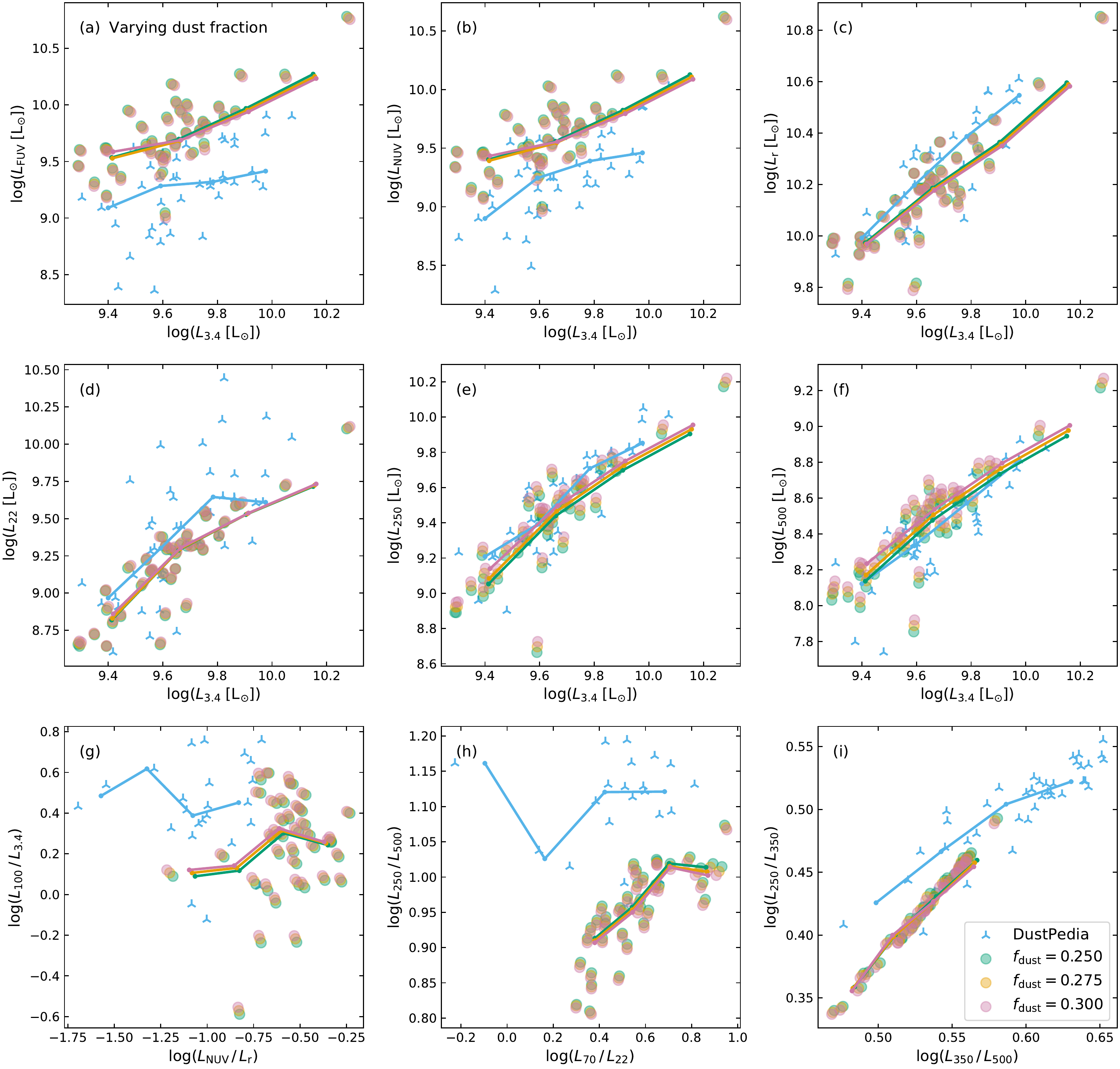}
  \caption{The same scaling relations as in Fig.~\ref{fig:scalingrelations}, now showing ARTEMIS luminosities calculated for a random viewing angle using the \sK21/\dT12 recipe with three values for the dust fraction $f_\mathrm{dust}=0.250$ (green), $0.275$ (orange), and $0.300$ (purple) near or at the optimal value (see Table~\ref{tab:fdust}) in comparison with observed DustPedia luminosities (blue).}
  \label{fig:scalingrelationsdustfraction}
\end{figure*}

As described in Sect.~\ref{sec:dustrecipes}, the recipes for handling diffuse dust in our RT simulations have a free parameter, the dust-to-metal fraction $f_\mathrm{dust}$, which must be determined through comparison with observations. We have chosen the luminosity scaling relations shown in Fig.~\ref{fig:scalingrelations} to accomplish this comparison, because they trace the key physical galaxy properties including stellar mass, SFR, sSFR, dust mass, and dust temperature. In principle, selecting an `optimal' $f_\mathrm{dust}$ value is a straightforward optimisation process. In practice, however, it is substantially complicated by several factors.

The numerical error on the ARTEMIS luminosities caused by discretization effects in the RT simulation is below 8 per cent or 0.033 dex (see Sect.~\ref{sec:calibrationconvergence}). The calibration error on the DustPedia fluxes for the broadbands used in our scaling relations is of the same order \citep[see Table 1 of][]{Clark2018}, although these numbers probably do not capture all observational uncertainties. For our purposes, we can assume that these variations constitute a minor factor.

Another factor is the effect of the sight line on the observed luminosities. While the DustPedia galaxies are obviously observed at some fixed sight line, we can control the viewing angle for our simulated ARTEMIS galaxies. The effect on our scaling relations is shown in Fig.~\ref{fig:scalingrelationssightline}. As expected, there is a significant discrepancy between the face-on and edge-on luminosities at shorter wavelengths, up to 0.37 dex in the $r$ band and up to 0.48 dex in the FUV band, while the effect is minimal at longer wavelengths. We mitigate the sight line factor by using a random viewing angle for the ARTEMIS galaxies in our calibration process, which probably corresponds most closely to the observed data set.

We need to quantify how well a given set of ARTEMIS data points corresponds to the observed DustPedia data points in the scaling relations. Following \citet{Camps2016} and \citet{Kapoor2021}, we employ a generalisation of the Kolmogorov-Smirnov test \citep[K--S test,][]{Kolmogorov1933,Smirnov1948} to two-dimensional distributions \citep{Peacock1983, Fasano1987, Press2002}. The 2D K--S test computes a metric which can be interpreted as a measure of the `distance' between two sets of two-dimensional data points. The metric may not be perfectly suited for our purposes because, for example, a shift away from the scaling relation is generally penalised in the same way as a more acceptable shift along the scaling relation. More importantly, to obtain a single overall measure for a given recipe, the metric for each of the relations must be aggregated. The final measure, and thus the ranking of different recipes, depends on the selection of scaling relations included in the metric and on the relative weights assigned to them.

As an example that is representative for the three recipes described in Sect.~\ref{sec:combinedrecipes}, Fig.~\ref{fig:scalingrelationsdustfraction} shows our scaling relations for results using the \sK21/\dT12 recipe with different dust fraction values: $f_\mathrm{dust}=0.250$, $0.275$, and $0.300$. The shifts in the scaling relations are much smaller than those shown for different sight lines in Fig.~\ref{fig:scalingrelationssightline}. Also, the shift can be toward or away from the observed DustPedia relation depending on the wavelength regime. For example, increasing the dust fraction and thus the attenuation at shorter wavelengths improves the tension with observations in the UV (panels a and b) but worsens it in the optical (panel c). Similarly, a shift towards colder dust temperatures (panel i) causes opposing shifts in the infrared relations (e.g., panel e versus panel f). Any aggregated metric based on these relations therefore depends significantly on the employed relative weights.

We experimented with various averaging schemes, always mixing results in UV, optical, and infrared wavelength regimes, and came to the following conclusions (see Table~\ref{tab:fdust}). For the \sC16/\dC16 recipe, $f_\mathrm{dust}=0.300$ is a robust optimal value, i.e. it comes out on top regardless of the averaging scheme. This corresponds to the optimal value found by \citet{Camps2016} for the EAGLE simulations, which is comforting because both the hydrodynamical simulation and RT post-processing recipes are very similar. By the same token, for the \sC16/\dT12 recipe, $f_\mathrm{dust}=0.275$ is a robust optimal value. The decrease in dust-to-metal ratio compared to the \sC16/\dC16 recipe can be understood by noting that the \dT12 scheme assigns dust to a broader set of gas particles than the \dC16 scheme (see Sect.~\ref{sec:dustrecipes}). For the ARTEMIS galaxies, this results in a `dusty' ISM mass that is 5 to 15 per cent higher. This increase is roughly compensated by the 9 per cent decrease in dust fraction. We do note, however, that the emitted radiation will also vary between the two recipes because of the changed relative dust/star geometry.

For the \sK21/\dT12 recipe, the values $f_\mathrm{dust}=0.275$ and $0.300$ result in very comparable statistics so that the `optimal' value sensitively depends on the chosen averaging scheme. We take this to mean that the actual optimal value lies between $0.275$ and $0.300$. This slight increase from the \sC16/\dT12 recipe must be related to the different handling of SF regions, which affects the dust-related emission modelled as part of the MAPPINGS III SED templates (see Sects.~\ref{sec:commonrecipes} and \ref{sec:sfrrecipes}). We choose to employ $f_\mathrm{dust}=0.275$ as the `optimal' value because the differences are small (see Fig.~\ref{fig:scalingrelationsdustfraction}) and we then have a consistent value for the \dT12 dust allocation scheme regardless of the recipe for handling SF regions. Also, \citet{Kapoor2021} find a fairly large difference in optimal dust fraction between the two dust allocation schemes (see Table~\ref{tab:fdust}).


\bsp	
\label{lastpage}
\end{document}